\documentclass[aps,pre,twocolumn,showpacs,notitlepage,floatfix,superscriptaddress]{revtex4-1}

\usepackage{amsmath}
\usepackage{amssymb}
\usepackage{mathtools}
\usepackage[all]{xy}
\usepackage{mathrsfs}
\input xy
\xyoption{matrix}
\xyoption{2cell}
\xyoption{arrow}
\xyoption{curve}
\xyoption{all}

\usepackage{graphicx}
\usepackage{epsfig}
\usepackage{dcolumn}
\usepackage{bm}

\usepackage{amsmath}
\usepackage{graphicx}
\usepackage[latin1]{inputenc}
\usepackage{ulem}
\usepackage{epstopdf}
\usepackage{subfigure}
\usepackage{color}

\RequirePackage[usenames,dvipsnames]{xcolor}
\usepackage{soul}

\begin{document}
\title{
Dynamical glass in weakly non-integrable Klein-Gordon chains
}

\author{Carlo Danieli}
\affiliation{Center for Theoretical Physics of Complex Systems, Institute for Basic Science, Daejeon, Korea}
\author{Thudiyangal Mithun}
\affiliation{Center for Theoretical Physics of Complex Systems, Institute for Basic Science, Daejeon, Korea}
\author{Yagmur Kati}%
\affiliation{Center for Theoretical Physics of Complex Systems, Institute for Basic Science, Daejeon, Korea}
\affiliation{Basic Science Program, Korea University of Science and Technology (UST), Daejeon 34113, Republic of Korea}
\author{David K. Campbell}
\affiliation{Boston University, Department of Physics, Boston, Massachusetts 02215, USA}
\author{Sergej Flach}
\affiliation{Center for Theoretical Physics of Complex Systems, Institute for Basic Science, Daejeon, Korea}
\affiliation{New Zealand Institute for Advanced Study, Massey University, Auckland, New Zealand}


\date{\today}

\begin{abstract}

Integrable many-body systems are characterized by a complete set of preserved actions. Close to an integrable limit, a {\it non-integrable} perturbation creates a coupling network in action space which can be short- or long-ranged. 
We analyze the dynamics of  observables which become the conserved actions in the integrable limit.
We compute distributions of their finite time averages and obtain the ergodization time scale $T_E$ on which these distributions converge to $\delta$-distributions.
We relate $T_E$ to the statistics of fluctuation times of the observables, which acquire fat-tailed distributions 
with standard deviations $\sigma_\tau^+$ dominating the means $\mu_\tau^+$ and establish that $T_E \sim (\sigma_\tau^+)^2/\mu_\tau^+$.  The Lyapunov time $T_{\Lambda}$ (the inverse of the largest Lyapunov exponent) is then compared to the above time scales.
We use a simple Klein-Gordon chain to emulate long- and short-range coupling networks by tuning its energy density.
For long-range coupling networks $T_{\Lambda}\approx \sigma_\tau^+$, 
which indicates that the Lyapunov time sets the ergodization time, with chaos quickly diffusing through the coupling network.
For short-range coupling networks we observe a {\it dynamical glass}, where  
$T_E$ grows dramatically by many orders of magnitude and greatly exceeds  the Lyapunov time, which satisfies $T_{\Lambda} \lesssim \mu_\tau^+$. 
This effect arises from the formation of highly fragmented inhomogeneous distributions of chaotic groups of actions, separated by growing volumes of non-chaotic regions. These structures persist up to the ergodization time  $T_E$.

\end{abstract}

\maketitle



\section{Introduction}

Ergodicity and mixing are central concepts in statistical mechanics.   
Both properties characterize the temporal evolution of a dynamical system:
ergodicity demands that a solution visits almost all states of the available phase space; mixing requires that as the system evolves any choice of two open sets of available states will eventually overlap \cite{lichtenberg1992regular}. 
With mixing being a necessary condition for ergodicity, both constitute fundamental aspects for the phenomenon of thermalization.
In particular, both properties imply that the available phase space does not fragment into inaccessible open sets, 
and that the infinite time average of any observable matches its phase space average.
The latter feature typically acts as a definition of ergodicity. 
The search for the violation of ergodicity, mixing and thermalization gave rise to a some of the most important discoveries in mathematics and statistical physics. 
Several of these results were found for {\it weakly non-integrable systems}:   
models of $N$ degrees of freedom whose Hamiltonians $H= H_0 + \bar \varepsilon H_1$ consist of an integrable part  $H_0$, a non-integrable perturbation $H_1$, and the perturbation strength $\bar \varepsilon$. 
In 1954 Kolmogorov proved the existence for small enough perturbation strength $0<\bar \varepsilon< \bar \varepsilon_0$ of sets with non-zero measure of infinite time stable solutions which are confined to $N$-dimensional manifolds of the phase space (later labeled {\it KAM tori}) \cite{Kolmogorov1954conservation}.
His work was extended by Arnold \cite{arnold1963aproof} and Moser \cite{Moser1962invariant} to larger classes of Hamiltonian systems, leading to the celebrated KAM theorem. 
Almost concomitantly with the discovery made by Kolmogorov, a numerical test on a small chain of harmonic oscillators in the presence of weak anharmonic coupling failed to show the expected equipartition of energy along the chain (original report \cite{Fermi:1955}, reviews in \cite{Ford:1992,gallavotti2007fermi,Porter:2009}). 
This computer experiment, performed by Fermi, Pasta, Ulam and Tsingou (FPUT), 
and the attempts to explain its striking outcomes, led to the discovery of solitons \cite{Zabusky:1965,Zabusky:1967} and to remarkable advances in the theory of Hamiltonian chaos \cite{Izrailev:1966, Chirikov1973numerical,Chirikov1979universal}. 
In particular, it was found that a large number of weakly non-integrable lattices possess families of exact time periodic solutions whose actions turn into the
ones of the integrable system $H_0$ for $\epsilon=0$. Depending on the limit being considered, these solutions are have been called  {\it discrete breathers} 
\cite{Mackay:1994proof,Flach:1998discrete,Campbell:2004localizing,Flach:2008discrete} and 
show typically exponential localization of energy in real space; or {\it q-breathers} which show exponential localization of energy in normal mode space
\cite{Flach:2005qBreathers,Flach:2006qBreathers}.
Although forming a set of zero measure, 
these coherent solutions can impact the dynamical properties of a many degree-of-freedom system, since a generic trajectory can spend long times in their neighborhoods in phase space\cite{tsironis1996slow,Rasmussen2000statistical,eleftheriou2003breathers,eleftheriou2005discrete,gershgorin2005renormalized,matsuyama2015multistage,zhang2016dynamical}.
These events (also labeled {\it excursions out of equilibrium}) have been experimentally studied in the context of semiconductor lasers \cite{Bonatto2011deterministic}, superfluids \cite{Ganshin2009energy}, microwave cavities \cite{Hohmann2010freak}, optical fibers \cite{Solli2007optical} and arrays of waveguides \cite{Eisenberg1998discrete,Lederer2008discrete}, among others.  

The impact of these excursions on the ergodic properties have been studied in the past 
for spin-glasses \cite{Bouchaud:1992weak} and time-continuous random walks  \cite{Bel2005weak,Bel2006random,Rebenshtok2007distribution,Rebenshtok2008weakly}. 
Recently an efficient numerical method to quantify the impact of the above excursions out of equilibrium in weakly non-integrable Hamiltonian systems has been proposed in Refs.\cite{Danieli2017Intermittent,Mithun2018weakly}.  
That scheme chooses the actions at the integrable limit as the relevant observables, and subsequently tracks their temporal
fluctuations. The resulting distributions of fluctuation times and their finite time average distributions 
permit one to extract a novel ergodization timescale $T_E$ \cite{Mithun2018JJ}. The dependence of $T_E$ on the strength of non-integrable perturbation $\bar \varepsilon$,
in particular its divergence for vanishing $\bar \varepsilon$ will signal the approaching of the integrable limit.

In this work we show that a non-integrable perturbation $H_1$ can span different classes of interaction networks among the actions of the integrable limit $H_0$.
These classes differ by network type being long range (respectively, short range). This distinct allows us to show that the Lyapunov time $T_\Lambda$ - the inverse largest Lyapunov exponent -
controls the ergodization dynamics and time scales for long range networks, whereas it does not in the short range case. We will use a simple
model - the Klein-Gordon (KG) chain for our computational studies. This model - as well as many other systems  - exhibits all the above qualitatively different
integrable limits.
The paper is organized as follows.  
In the next section, we introduce weakly non-integrable Hamiltonian systems and define long and short networks of actions. 
We then present our numerical studies of the different integrable limits of
the Klein-Gordon chain. In the concluding section, we recap and discuss our results. A series of Appendices provide technical details.

 \section{Weakly non-integrable Hamiltonian systems and networks of actions}
\label{section2}

Consider a Hamiltonian $H$ with $N$ degrees of freedom 
\begin{equation}
H = H(p,q)
\label{eq:ham1}
\end{equation}
where $q = (q_1,\dots,q_N)$ are the position coordinates and $p = (p_1,\dots,p_N)$ are the conjugate momenta. 
These coordinates belong to the $2N$ dimensional phase space $X = \mathbb{R}^N\times \mathbb{R}^N$. 
The equations of motion are
\begin{equation}
\dot{p}_n = - \frac{\partial H}{\partial q_n}\ ,\qquad 
\dot{q}_n= \frac{\partial H}{\partial p_n} \;.
\label{eq:ham_eq1}
\end{equation}
An integral of motion $I$ ({\it e.g.} the Hamiltonian energy $H$) is a quantity that is conserved along the solutions of Eq.(\ref{eq:ham_eq1}). 
The existence of $\ell$ integrals of motions implies that 
a trajectory is confined to a co-dimension $\ell$ submanifold, called the {\it available phase space}.
A Hamiltonian $H$ is called {\it integrable} if there exists a canonical transformation $(p,q) = \phi(J,\theta) $ that maps the conjugate coordinates $(p,q)$ into action-angle coordinates $ (J,\theta)$ such that
\begin{equation}
 H(\phi(J,\theta)) \equiv H_0(J) 
\label{eq:ham_tr1}
\end{equation}
so that the Hamiltonian $H_0$ depends only on the actions $ \{J_n\}_{n=1}^N$.
The existence of such a canonical transformation $\phi$ is ensured by the {\it Liouville-Arnold theorem} \cite{Arnold1988Mathematical}.
The equations of motion Eq.(\ref{eq:ham_eq1}) of an integrable system expressed in action-angles coordinates read
\begin{equation}
\begin{split}
\dot{J}_n  = - \frac{\partial H_0}{\partial \theta_n} = 0  \;, \qquad 
\dot{\theta}_n & = \frac{\partial H_0}{\partial J_n}=  \omega_n(J) 
 \end{split} \;.
\label{eq:ham_eq_can_var1}
\end{equation}
Solutions of Eq.(\ref{eq:ham_eq_can_var1}) yield constant actions $J_n(t)$ and time-periodic angles $\theta_n(t)$ that wind on $N$-dimensional tori $\mathbb{T}^N$.
\begin{equation}
\begin{split}
&J_n(t) = J_n^0  \;,  \qquad  \theta_n(t) = \omega_n t + \theta_n^0
 \end{split}
\label{eq:ham_eq_IL_col}
\end{equation}
for the frequencies $\omega_n$.
Consequently, the phase space $X$ is foliated by a set of invariant tori $\mathbb{T}^N$, where the solutions in Eq.(\ref{eq:ham_eq_IL_col}) are confined for all times $t\in \mathbb{R}$. 

Let us consider a general Hamiltonian $H$ and define the energy density $h=H/N$.  
If in regimes of small or large $h$ some of its terms become negligible with respect to an otherwise integrable reminder $H_0$, we say that $H$ possesses an {\it integrable limit}. 
This can be realized by considering a Hamiltonian of the form 
\begin{equation}
H = H_0 + \bar \varepsilon H_1\label{eq:perturbed_ham1}
\end{equation}
where $H_0$ is integrable, and $H_1$ is a non-integrable perturbation whose strength is controlled by a small parameter $0< \bar \varepsilon\ll1$. 
Then Eq.(\ref{eq:ham_eq1}) reads 
\begin{equation}
\begin{split}
\dot{\theta}_n & =  \omega_n(J) + \bar \varepsilon V_n(J,\theta)\ ,\qquad  
\dot{J}_n  = -\bar \varepsilon W_n(J,\theta)
\end{split}
\label{eq:ham_eq_can_var2}
\end{equation}
where $V_n = \partial H_1/\partial J_n$ and $W_n = \partial H_1/\partial \theta_n$. 
We shall call the system in Eq.(\ref{eq:perturbed_ham1}) 
weakly non-integrable. 
For $\bar \varepsilon\neq 0$, 
the term $W_n$ links the time-dynamics of an action $J_n$ to all or a subset of actions and angles of the system. 
In a typical case,
each action $J_k$ is connected to a number $R_n$ of
groups of actions $\{ G_m\}_{m=1}^{R_n}$, each one formed by $L_{n,m}$ actions 
$G_m = \{J_{g_{n,m}(l)}\}_{l=1}^{L_{n,m}}$. 
It then follows that 
a non-integrable perturbation defines a network between the actions  $ \{J_n\}_{n=1}^N$, where $R_n$ and $L_{n,m}$ depend on $H_1$. 
We henceforth
distinguish networks according to how the number of groups of actions linked by the perturbation $H_1$ depends on the number of degrees of freedom $N$.
Let us define the {\it coupling range} $\mathcal{R} = max\{R_n | n\leq N\}$. We can then distinguish the following two cases: \\
{\bf Long Range Network (LRN)}: the coupling range $\mathcal{R}$ scales with the number $N$ of degrees of freedom of the system $\mathcal{R} = g(N)$ for a certain monotonic function $g$; \\
{\bf Short Range Network (SRN)}: the coupling range $\mathcal{R}$ is finite and independent from the number $N$ of degrees of freedom of the system.

\section{The model}

We consider a class of classical translation invariant interacting many-body systems described by the Hamiltonian
\begin{equation}
\label{eq:ham_gen}
H = \sum_{n=1}^{N}\left[ \frac{p_n^2}{2} + V(q_n) + \varepsilon W(q_{n+1} - q_n)\right] 
\end{equation} 
where  $V$ is a local potential and $W$ is an interaction potential, with $V(0)=W(0)= V^\prime(0) = W^\prime(0)=0$, $V^{\prime \prime}(0) , W^{\prime \prime}(0)>0$.
We focus on the Klein-Gordon (KG) chain, for which
\begin{equation}
\label{eq:KG_2}
V(q)=\frac{q^2}{2} + \frac{q^4}{4}\ ,\quad W(q)= \frac{q^2}{2} \ .
\end{equation}  
The equations of motion Eq.(\ref{eq:ham_eq1}) read
\begin{align}
\label{eq:KG_equation2}
\ddot{q}_n &= - q_n - q_n^3 + \varepsilon(q_{n+1} + q_{n-1} - 2q_{n}) \ .
\end{align}  
Let us discuss several different integrable limits using $\varepsilon$ and the energy density $h$ as control parameters.
For $h=const, \varepsilon \rightarrow 0$ and equally $h \rightarrow \infty, \varepsilon = const$ the system reaches an integrable set of
decoupled oscillators,
with the interaction potential $W$ acting as the perturbation $H_1$ in Eq.(\ref{eq:perturbed_ham1}), 
and the network is SRN.
In contrast, $h\rightarrow 0, \varepsilon = const$ is a LRN integrable limit, since the quartic term in the local potential $V$ 
becomes negligible with respect to the remaining quadratic ones. 
In this limit, the KG chain reduces to a chain of harmonic oscillators. The term $q_n^4 / 4$ (transformed to Fourier space) couples all the normal modes to each other
and yields a long range network.
We note that the KG Hamiltonian possesses only one conserved quantity $H$.
Eq.(\ref{eq:KG_equation2}) will be integrated in time using symplectic schemes (see appendix \ref{app:numerical}). 
\

\section{ Methods}

We follow the dynamics of time-dependent observables $J$ which become conserved at the chosen integrable limit.
{\it Ergodicity} away from the integrable limit would imply that their {\it infinite} time average equals their statistical average $\langle J\rangle$
\begin{equation}
\lim_{T\rightarrow\infty} \overline{J}_T=  
\lim_{T\rightarrow\infty}   = \frac{1}{T}\int_0^T J(t)dt = 
\langle J\rangle \ .
\label{eq:ergodicity}
\end{equation}  
We numerically compute the finite time averages $\overline{J}_{T}$ for $M$ different trajectories and obtain their distribution $\rho(\overline{J};T)$. 
This distribution is 
characterized by a first moment $\mu_J$ and  a non-zero variance $\sigma_J^2(T)$. 
From Eq.(\ref{eq:ergodicity}) it follows that   
$\rho(\overline{J};T\rightarrow \infty) = \delta(\overline{J} - \langle J\rangle)$. 
We study this convergence by computing the dimensionless {\it squared coefficient of variation} 
(also called  {\it fluctuation index})
\begin{equation}
\label{eq:sigma_var}
q(T) = \frac{\sigma_J^2(T)}{ \mu_J^2} \ .
\end{equation}
Following \cite{Mithun2018JJ},  we extract an ergodization timescale $T_E$ as
\begin{equation}
q(T)\sim 
\begin{cases}
q(0) \quad &\text{for} \ T\ll T_E \\
\frac{T_E}{T}  &\text{for} \  T\gg T_E
\end{cases}
\label{eq:qT_TE}
\end{equation}
We further study the fluctuation statistics of $J(t)$  
by computing the {\it piercing times} $t^i$ 
at which $J(t) = \langle J\rangle$.
The {\it excursion times} 
\begin{equation}
\tau^\pm (i) = t^{i+1} - t^i
\label{eq:excursions}
\end{equation}
distinguish between $J(t) > \langle J \rangle$ ($\tau^+$) and $J(t) < \langle J \rangle$ ($\tau^-$). 
We then compute numerically the distributions $P^\pm $ of the excursion times $\tau^\pm$, as well as their averages $\mu_{\tau}^\pm$ and their standard deviations $\sigma_{\tau}^\pm$ (see appendix \ref{app:PDF} for details). 
In particular, we will focus on the $\tau^+$ events (see appendix \ref{sec:tau_pm}). 
These timescales dictate the ergodization timescales $T_E$ according to Ref. \cite{Mithun2018JJ}
\begin{equation}
T_E \sim 
\tau_{q}^+ \equiv 
\frac{ \big( \sigma_{\tau}^+ \big)^2}{\mu_{\tau}^+}\ .
\label{eq:qT_asymp}
\end{equation}
when neglecting correlations between different events
(see appendix \ref{sec:q_1/T} for details). 
Finally, we relate the ergodization time $T_E$ to the Lyapunov time
\begin{equation}
T_{\Lambda} = 1 / \Lambda \;
\label{eq:Lyapunov_time}
\end{equation}
defined as the inverse of the largest Lyapunov exponent  $\Lambda$ and numerically obtained via the tangent method (see appendix \ref{app:Lyapunov} for details).  
The dynamics of a non-integrable system will be essentially identical to that of an integrable (but usually unknown) approximation precisely up to the time scale $T_{\Lambda}$.

\section{Long range network}

Let us consider the small energy limit $h\rightarrow 0,\varepsilon=const$ of the KG chain. 
The integrable Hamiltonian $H_0$ consists of a chain of harmonic oscillators
\begin{equation}
H_0 = \sum_{n=1}^N \left[\frac{p_n^2}{2} +  \frac{q_n^2}{2} + \frac{\varepsilon}{2}(q_{n+1} - q_{n})^2 \right] \; .
\label{eq:KG_int_ham_E0}
\end{equation} 
The non-integrable perturbation is then given by
\begin{equation}
\bar{\epsilon} H_1 = \sum_{n=1}^N \frac{q_n^4}{4}\;.
\label{eq:KG_nonint_anharm_E1}
\end{equation}
We choose fixed boundary conditions $p_0=p_{N+1}=q_0=q_{N+1}=0$, in analogy with the small energy limit of the FPU chain \cite{Fermi:1955}, in order to remove
degeneracies of eigenmode frequencies.

We use the 
canonical transformation to normal mode momenta and coordinates $\{P_k,Q_k\}$
\begin{equation}
\begin{split}
\left(
\begin{array}{c}
P_k\\
Q_k\\
\end{array}
\right)
&= \sqrt{\frac{2}{N+1}}\sum_{n=1}^{N}
\left(
\begin{array}{c}
p_n\\
q_n\\
\end{array}
\right)
 \sin \bigg(\frac{\pi n k}{N+1}\bigg)
\end{split}
\label{eq:Fourier coordinates}
\end{equation}
for $k=1,\dots,N$. This transformation diagonalizes the Hamiltonian $H_0 = \sum_{k=1}^N E_k$ in Eq.(\ref{eq:KG_int_ham_E0}), 
where the normal mode energies $E_k$ are 
\begin{equation}
E_k = \frac{P_k^2 + \Omega_k^2 Q_k^2}{2}\ , \qquad \omega_k = 2\sin\left(\frac{\pi k }{2(N+1)}\right)
\label{eq:modes_energy_KG}
\end{equation}
for $\Omega_k \equiv \sqrt{1+\varepsilon \omega_k^2}$. 
The equations of motion 
Eq.(\ref{eq:KG_equation2}) in the normal mode coordinates Eq.(\ref{eq:Fourier coordinates}) then read 
\begin{equation}
\ddot{Q}_k + \Omega_k^2  Q_k =  -\frac{1}{2(N+1)}  \sum_{l_1,l_2,l_3}  A_{k,l_1,l_2,l_3} Q_{l_1} Q_{l_2} Q_{l_3} 
\label{eq:KG_NM_low_e}
\end{equation}
where
\begin{equation}
\begin{split}
A_{k,l_1,l_2,l_3} &= 
\delta_{ k - l_1 + l_2 - l_3 , 0}  + \delta_{k - l_1 - l_2 + l_3,0}  \\
&-  \delta_{k+ l_1+l_2 - l_3,0}  -  \delta_{k + l_1 - l_2  + l_3 , 0}
\end{split}
\label{eq:KG_exp_coeff1}
\end{equation}
represents the coupling between the Fourier coordinates $Q_k$. 
Using the canonical transformation  
\begin{equation}
\label{eq:can_transf_KG_e0}
Q_k = \sqrt{2 J_k} \sin \theta_k  \qquad P_k = \Omega_k \sqrt{2 J_k} \cos \theta_k 
\end{equation}
it follows
\begin{equation}
\label{eq:KG_action_angle}
\begin{split}
&\qquad  \dot{J}_k 
 = - \frac{1}{\Omega_k} \sum_{l_1,l_2,l_3} \mathscr{A}_{k,l_1,l_2,l_3}  \sqrt{J_k J_{l_1} J_{l_2}  J_{l_3}} 
\end{split}
\end{equation}
where the coefficients $\mathscr{A}_{k,l_1,l_2,l_3} $ depend on the angles $\{ \theta_k\}_k$
\begin{equation}
\label{eq:KG_action_angle_coeff}
\begin{split}
&\mathscr{A}_{k,l_1,l_2,l_3} 
 = \frac{ A_{k,l_1,l_2,l_3} }{2(N+1)} \cos \theta_k \sin \theta_{l_1} \sin \theta_{l_2} \sin \theta_{l_3} \ .
\end{split}
\end{equation}
For each action $J_k$, the sum in Eq.(\ref{eq:KG_action_angle})) involves $R_k =  N^2 $ groups of $L_k = 4$ variables $\{J_k\}_k$ and is constrained by the conditions implied by Eq.(\ref{eq:KG_exp_coeff1}).  
Hence the coupling range $\mathcal{R}$ scales with the number of degrees of freedom $N$, and this limit leads to 
a long range network of actions $J_k$, similarly to the FPU case discussed in \cite{Flach2008Periodic}.
\begin{figure}[h]
 \centering
 \includegraphics[ width=0.885\columnwidth]{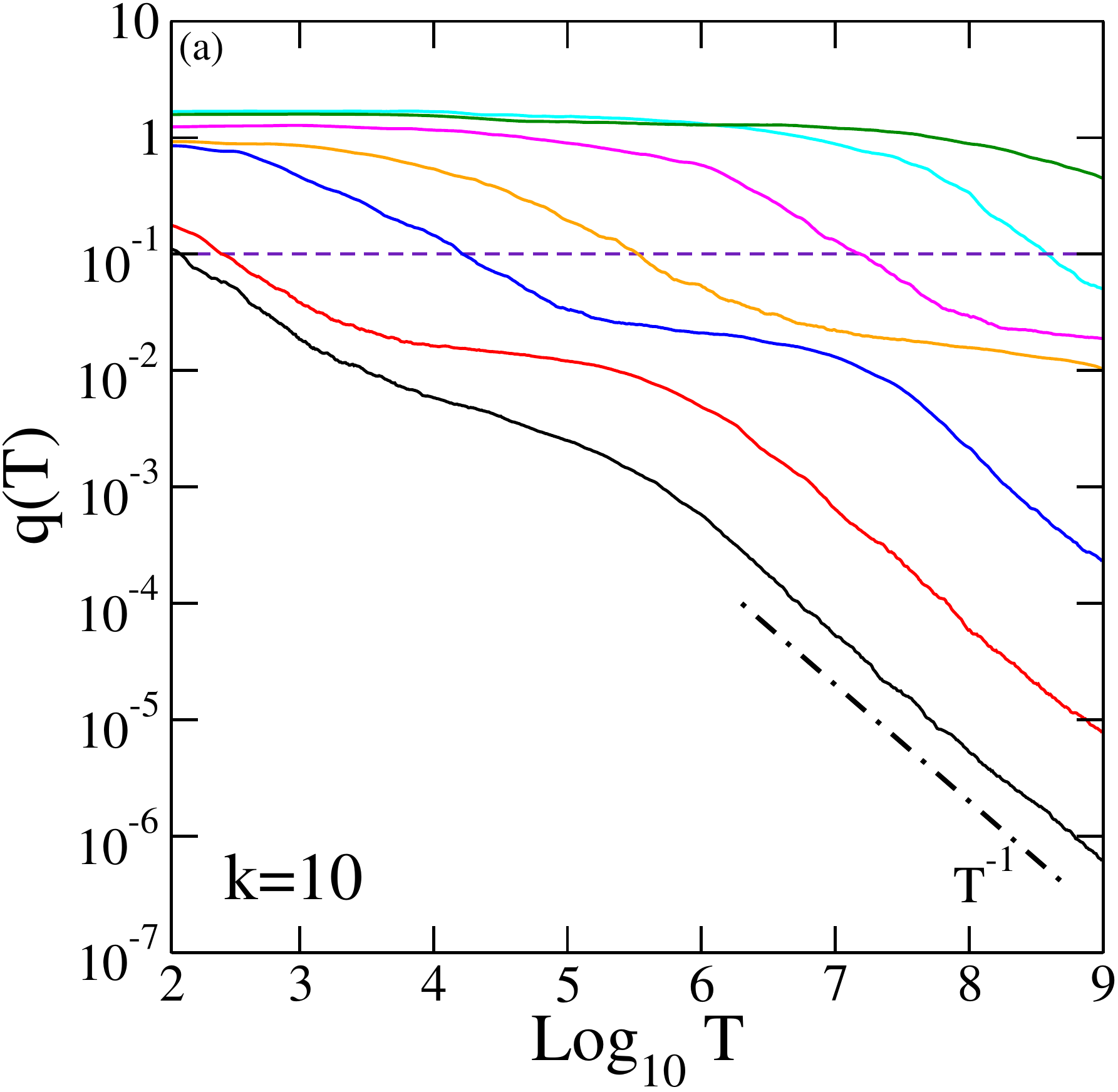}
 \includegraphics[ width=0.885\columnwidth]{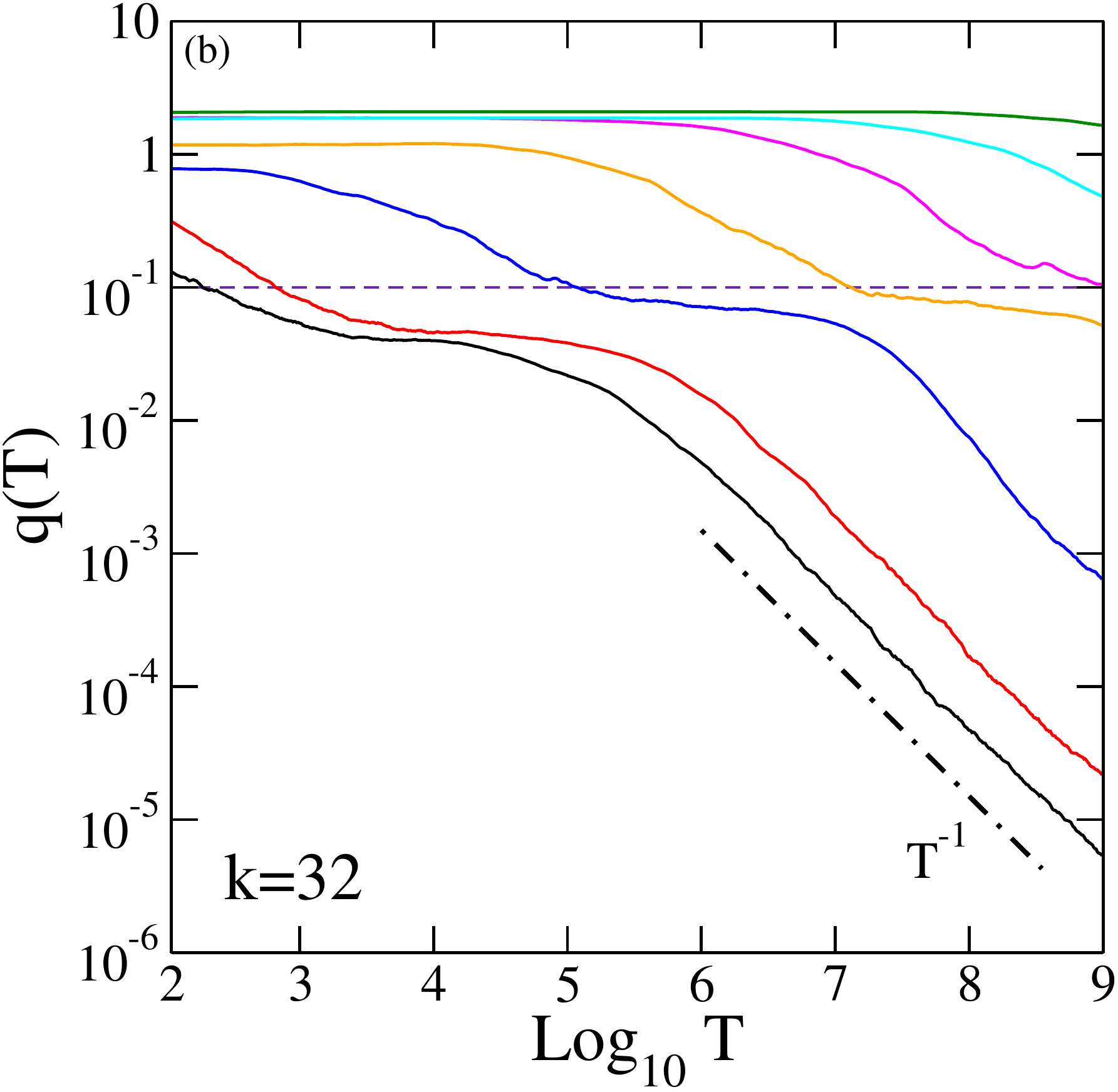}
 \caption{
(a) Squared coefficient of variation $q(T)$ 
computed for (top to bottom) 
$h=0.0025$ (green),
$h=0.005$ (cyan),
$h=0.01$ (magenta),
$h=0.025$ (orange), 
$h=0.1$ (blue),  
$h=1$ (red); 
$h=3.5$ (black)
obtained for mode $k=10$.
(b) Same as (a) for $k=32$.  
The black dashed-dotted lines guide the eye and indicate the algebraic decay.
The violet dashed horizontal lines indicate the $q=0.1$ threshold. 
Here $\varepsilon = 1$, $N=2^5$, $M=2^9$.
}
  \label{fig1}
\end{figure}
\begin{figure}[h]
 \centering
 \includegraphics[ width=0.9\columnwidth]{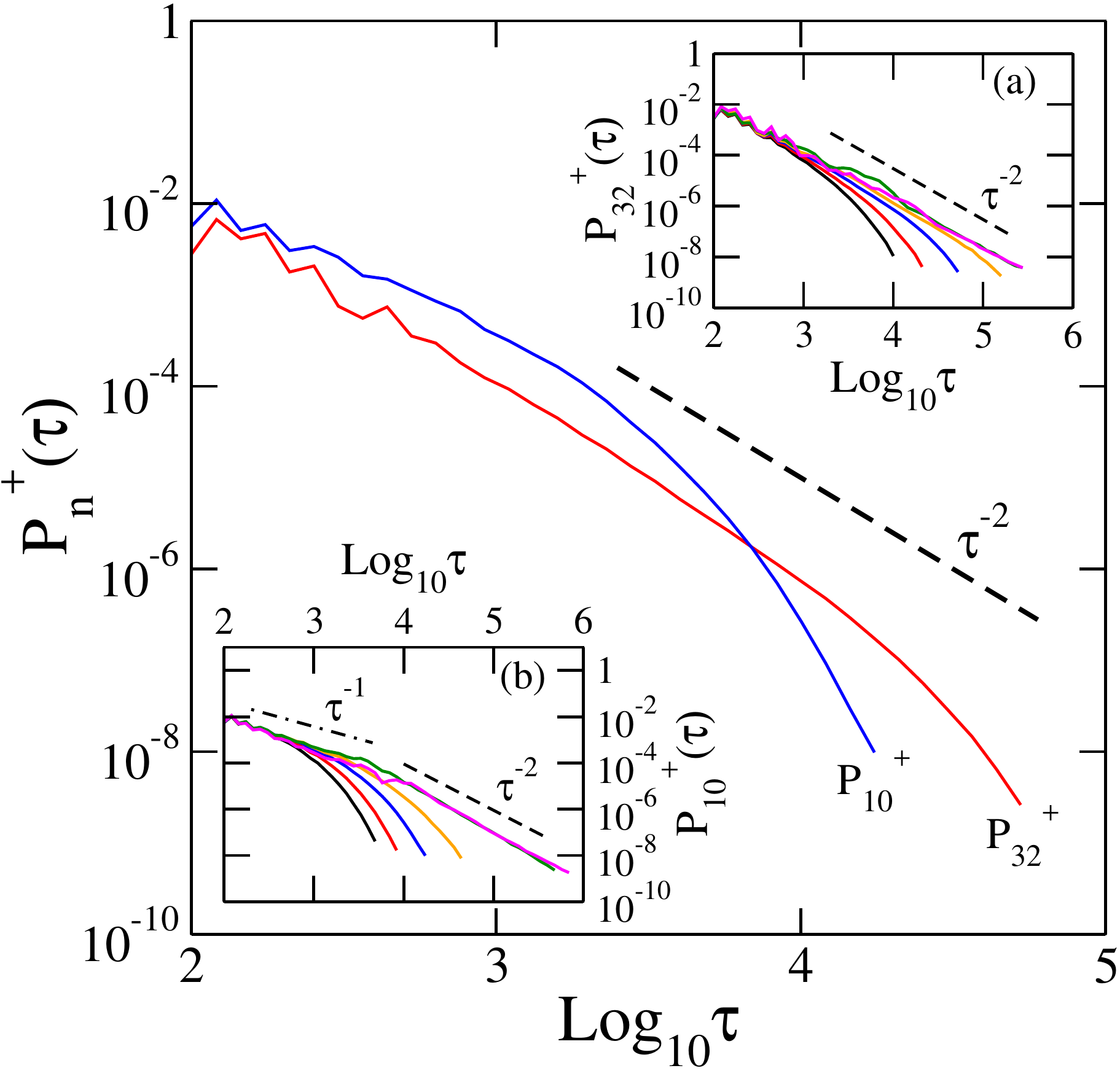}
 \caption{$P_{10}^+(\tau)$ (blue) and $P_{32}^+(\tau)$ (orange) for $h = 0.05$. 
Inset (a): $P_{32}^+(\tau)$ obtained for
$h=0.1$ (black), 
$h =0.075$ (red),
$h =0.05$ (blue),
$h =0.03$ (orange),
$h =0.01$ (green), and 
$h =0.005$ (magenta). 
 (b) Same as inset (a) for $P_{10}(\tau)$. 
The dashed lines guides the eye and indicate the algebraic decay. 
Here $N=2^5$, $\varepsilon = 1$, $ M=2^9$ and $T=10^{9}$. 
}
  \label{fig2}
\end{figure}
\begin{figure}[h]
 \centering
 \includegraphics[ width=0.9\columnwidth]{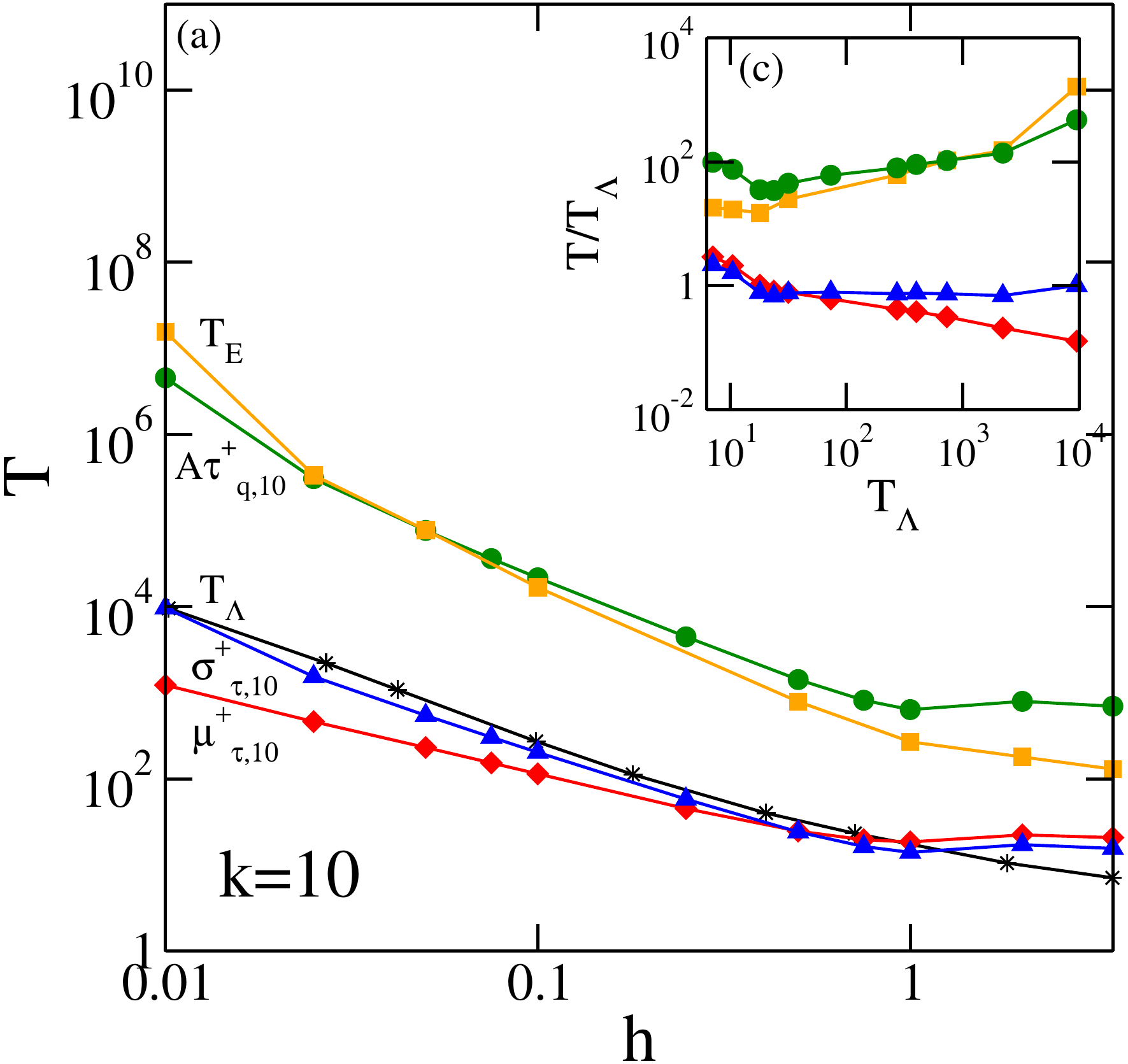}
 \includegraphics[ width=0.9\columnwidth]{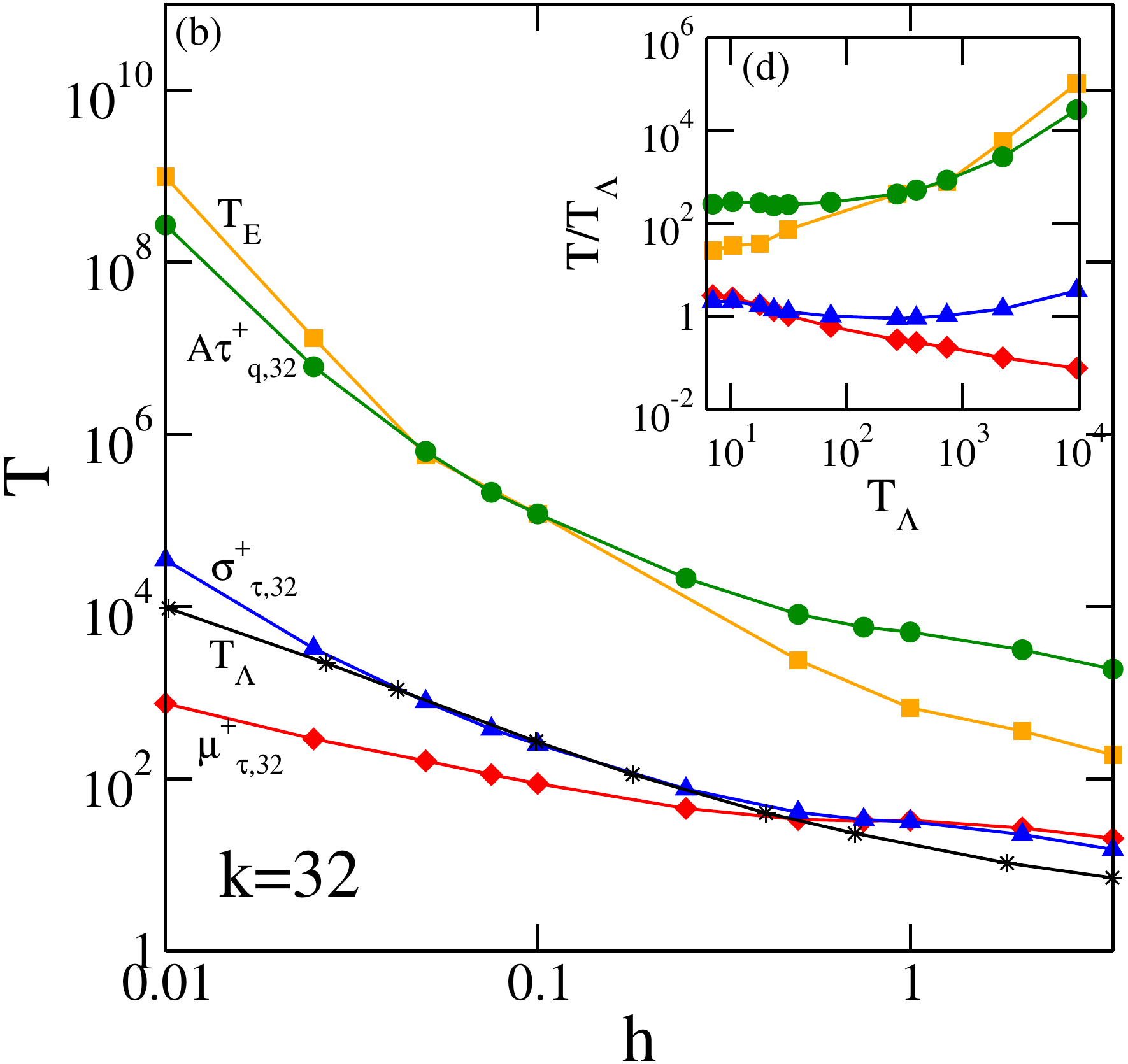}
 \caption{
 (a) Time scales $T_E$ (orange squares), $A\tau_{q,k}^+$ (green circles), $\sigma_{\tau,k}^+$ (blue triangles), $\mu_{\tau,k}^+$ (red diamonds)
and  $T_{\Lambda}$ (black stars) 
vs. the energy densities $h$ for $k=10$.
 (b) Same as (a) but for $k=32$. 
 Inset (c) rescaled times ($T_E$, $A\tau_{q,k}^+$, $\sigma_{\tau,k}^+$, $\mu_{\tau,k}^+$) in units of $T_{\Lambda}$.
 Here $N=2^5$, $\varepsilon = 1$, $M=2^9$, $A= 166$ and $T=10^{9}$. 
}
  \label{fig3}
\end{figure}

We use the normal mode energies $E_k = \Omega_k^2 J_k$ as the time-dependent observables, which are statistically distinguishable.
In Fig.~\ref{fig1}, we show $q(T)$ for two different modes: one located in the band center $k=10$ [plot(a)] and one in the band edge $k=32$ [plot(b)] for different $h$ fixing $\varepsilon = 1$ for a system of $N=2^5$ oscillators and averaging over $M=2^9$ initial conditions.
In both cases, the asymptotic decay $q(T)\sim T_E/T$ 
is visible for the larger energy cases (from black to blue).
We estimate $T_E$ using $q(T_E) = 0.1$ (horizontal dashed lines in Fig.~\ref{fig1} - see appendix \ref{sec:TE} for details).
For $q=0.1$ the distribution $\rho$ shows substantial  converge to its limiting delta function profile, and varying the cut-off condition does not affect the 
outcome up to a common scaling prefactor
(see appendix \ref{sec:distro_T_ave} for examples).

In Fig.~\ref{fig2} we show the distributions for the $k=32$ mode ($P_{32}^+$ (red)) and the $k=10$ mode ( $P_{10}^+$ (blue)) of the excursion times $\tau_k^+$ for $h = 0.05$ and $\varepsilon = 1$. 
These distributions differ from each other in accord with the statistical distinguishability of the actions.   
In the insets (a) and (b) of Fig.~\ref{fig2} we show $P_{32}^+$ and $P_{10}^+$ respectively for various values of $h$. Both cases show intermediate (though different) power-law tail trends: namely $P_{32}^+$ shows intermediate $\tau^{-2}$, while  $P_{10}^+$ shows two consecutive intermediate power-law transient regions $\tau^{-1}$ and $ \tau^{-2}$ respectively. 
We extract and use in the following the 
averages $\mu_{\tau,k}^+$ and the standard deviations $\sigma_{\tau,k}^+$ of the excursion times $\tau^+$.
.

In Fig.~\ref{fig3} we compare all computed time-scales for mode number $k=10$ [plot(a)] and $k=32$ [plot(b)] as a function of $h$. 
We plot the measured $T_E$ with orange squares, while we use red diamonds for averages $\mu_{\tau,10}^+,\mu_{\tau,32}^+$  and blue triangles for deviations $\sigma_{\tau,10}^+,\sigma_{\tau,32}^+$.
We observe that $\sigma_{\tau,k}^+ \approx \mu_{\tau,k}^+$ at $h=1$. For $h \rightarrow 0 $ $\sigma_{\tau,k}^+ \gg \mu_{\tau,k}^+$ in accord with the
above observed fat tails of the corresponding distribution functions $P$.
We then plot $A\tau_{q,k}^+$ using green circles for a fitting parameter $A=166$ (see \cite{ratio_A} for details) and confirm the predicted 
relation between the excursion times statistics and the ergodization time $T_E$ in Eq.(\ref{eq:qT_asymp}). 
Finally we plot 
in Fig.~\ref{fig3} the Lyapunov time $T_\Lambda$ (black stars). 
In both cases $k=10$ and $k=32$ $T_{\Lambda} \approx \sigma_{\tau,k}^+$, which indicates that the fat tails of the distributions of excursion times
are controlled 
by the Lyapunov time. 
To illustrate this, we show $T_E$, $\mu_{\tau,k}^+$, $\sigma_{\tau,k}^+$ and $A\tau_{q,k}^+$ in units of $T_{\Lambda}$, and as a function of $T_{\Lambda}$
in the insets of Fig.~\ref{fig3}.

\section{Short range network}

We use to periodic boundary conditions $p_1=p_{N+1}$, $q_1=q_{N+1}$.
In the limit of weak coupling $\varepsilon\ll 1,h=const$ (respectively $h\gg 1, \varepsilon = const$)
the system (\ref{eq:ham_gen}-\ref{eq:KG_2}) is close to an integrable limit with
an integrable Hamiltonian $H_0$ of a chain of decoupled anharmonic oscillators:
\begin{equation}
H_0 = \sum_{n=1}^N \left[\frac{p_n^2}{2} +  \frac{q_n^2}{2} + \frac{q_n^4}{4}  \right] \;.
\label{eq:KG_int_ham_E_inf}
\end{equation} 
The non-integrable perturbation is then given by
\begin{equation}
\bar{\epsilon} H_1= \frac{\varepsilon}{2}\sum_{n=1}^N (q_n-q_{n-1})^2\;.
\label{eq:KG_int_ham_E_inf_2}
\end{equation}
$H_1$ couples only nearest neighboring oscillators, 
leading to a SRN of actions.
As in \cite{Ivanchenko2004discrete}, we choose 
\begin{equation}
I_n = \frac{p_n^2}{2} + V(q_n) + \frac{\varepsilon}{4}\big[ (q_{n+1} - q_n)^2  + (q_{n} - q_{n-1})^2 \big]
\label{eq:KG_observables}
\end{equation}
as the time-dependent observables, which become conserved in both integrable limits.
Due to translation invariance, the observables $I_n$ are statistically equivalent, fluctuating around the energy density $h$. 
Since their distributions of finite time averages and of fluctuation times are identical, we extract measurements from all sites
and use them for the computation of the distributions (see appendix \ref{sec:PDF_KG_largeE}). That allows us to reduce the number of trajectories studied.
\begin{figure}[h]  
 \centering
\includegraphics[ width=0.9\columnwidth]{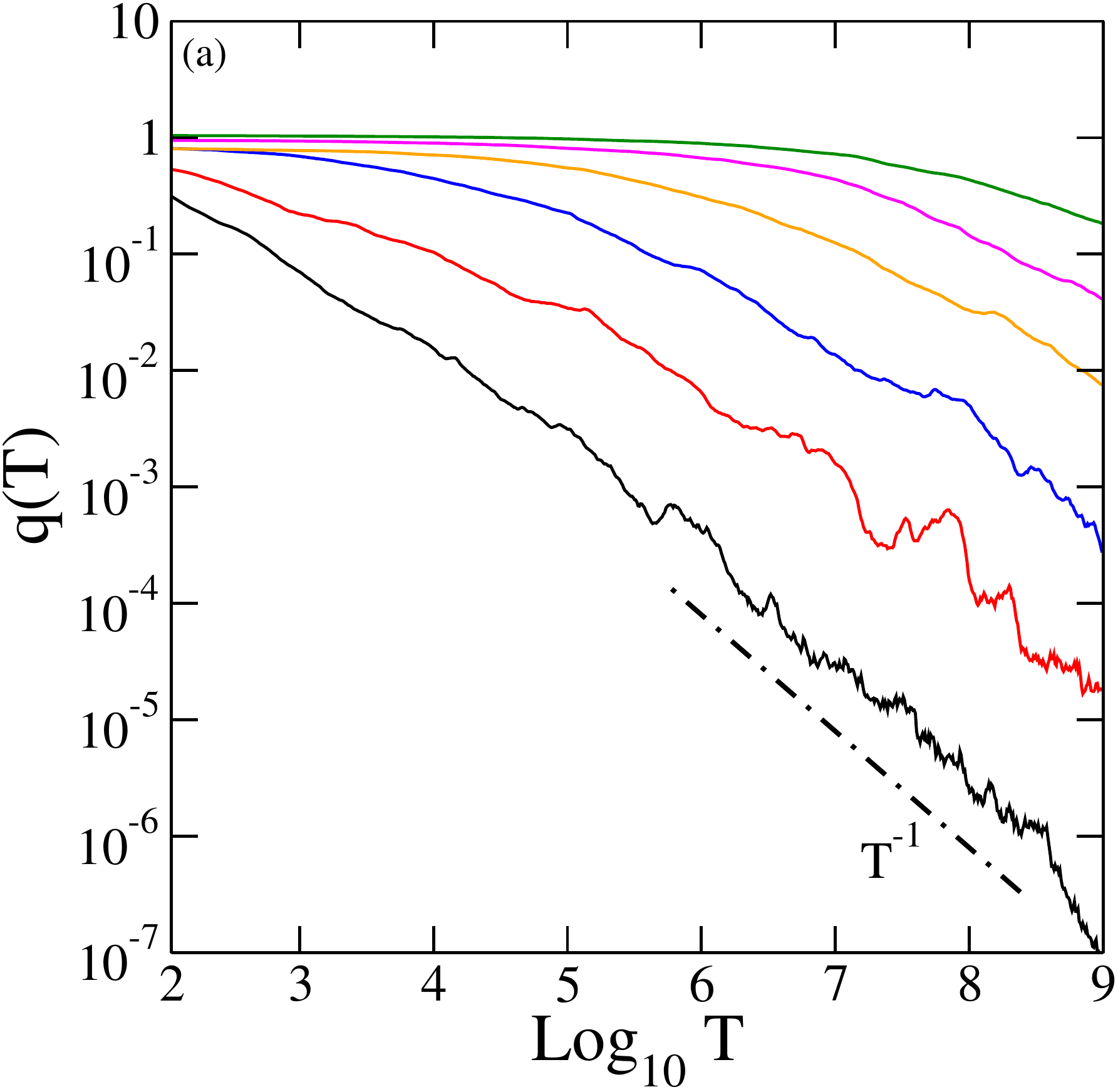}
\includegraphics[ width=0.9\columnwidth]{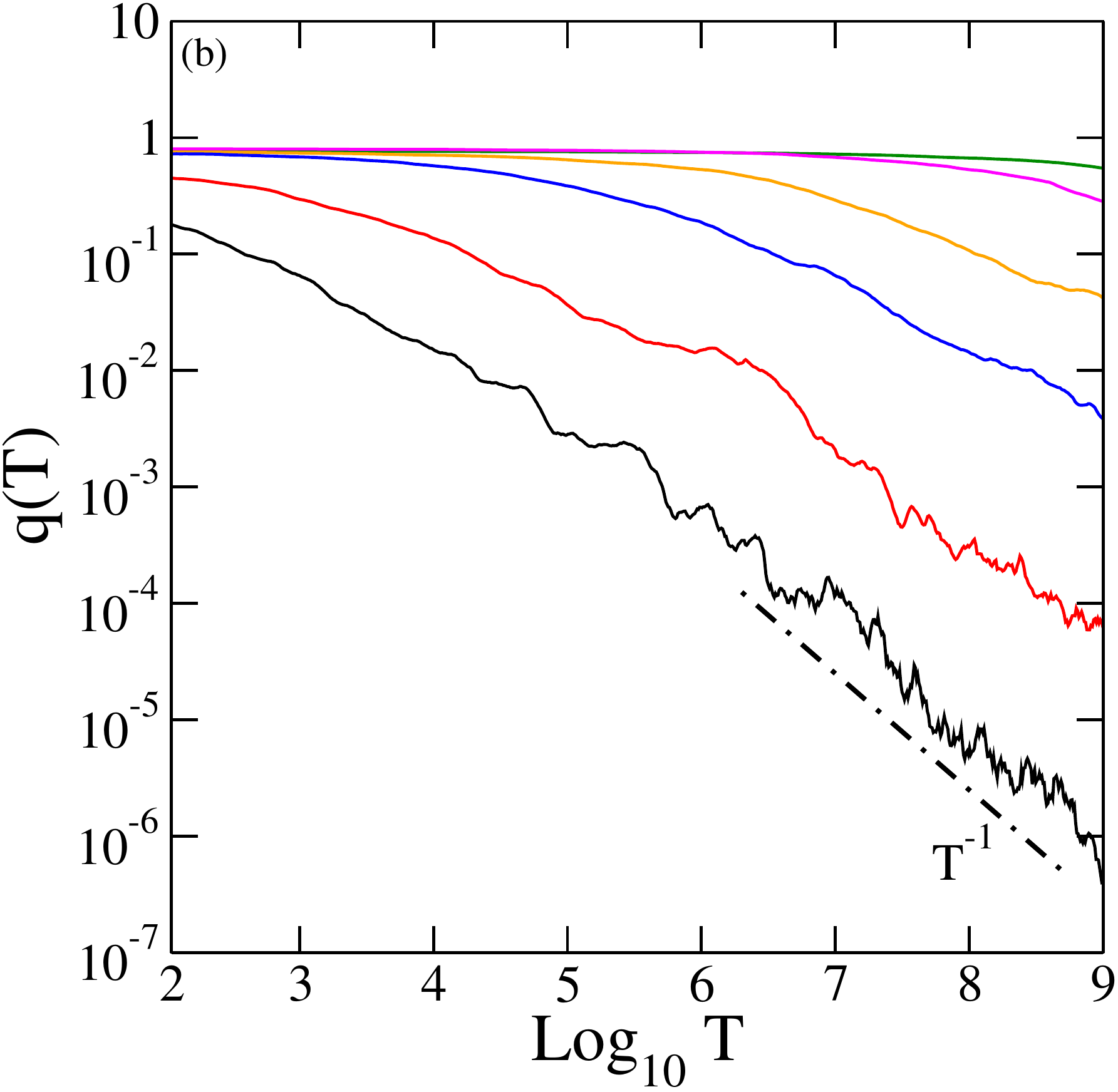}
 \caption{
(a) Squared coefficient of variation $q(T)$ 
computed for (top to bottom) 
$h=12$ (green),
$h = 6$ (magenta),
$h = 3$ (orange),
$h = 0.5$ (blue), 
$h = 0.1$ (red), 
$h = 0.01$ (black) 
with $\varepsilon = 0.05$.
(b) Same as (a) for (top to bottom)
$\varepsilon=0.015$ (green),
$\varepsilon=0.025$ (magenta), 
$\varepsilon=0.05$ (orange), 
$\varepsilon=0.1$ (blue),  
$\varepsilon=0.3$ (red), 
$\varepsilon=0.8$ (black)  
with $h = 5$.  
The black dashed-dotted lines guide the eye and indicate the algebraic decay. 
Here $N=2^{10}$
}
  \label{fig4}
\end{figure}
\begin{figure}[h]
 \centering
 \includegraphics[ width=0.9\columnwidth]{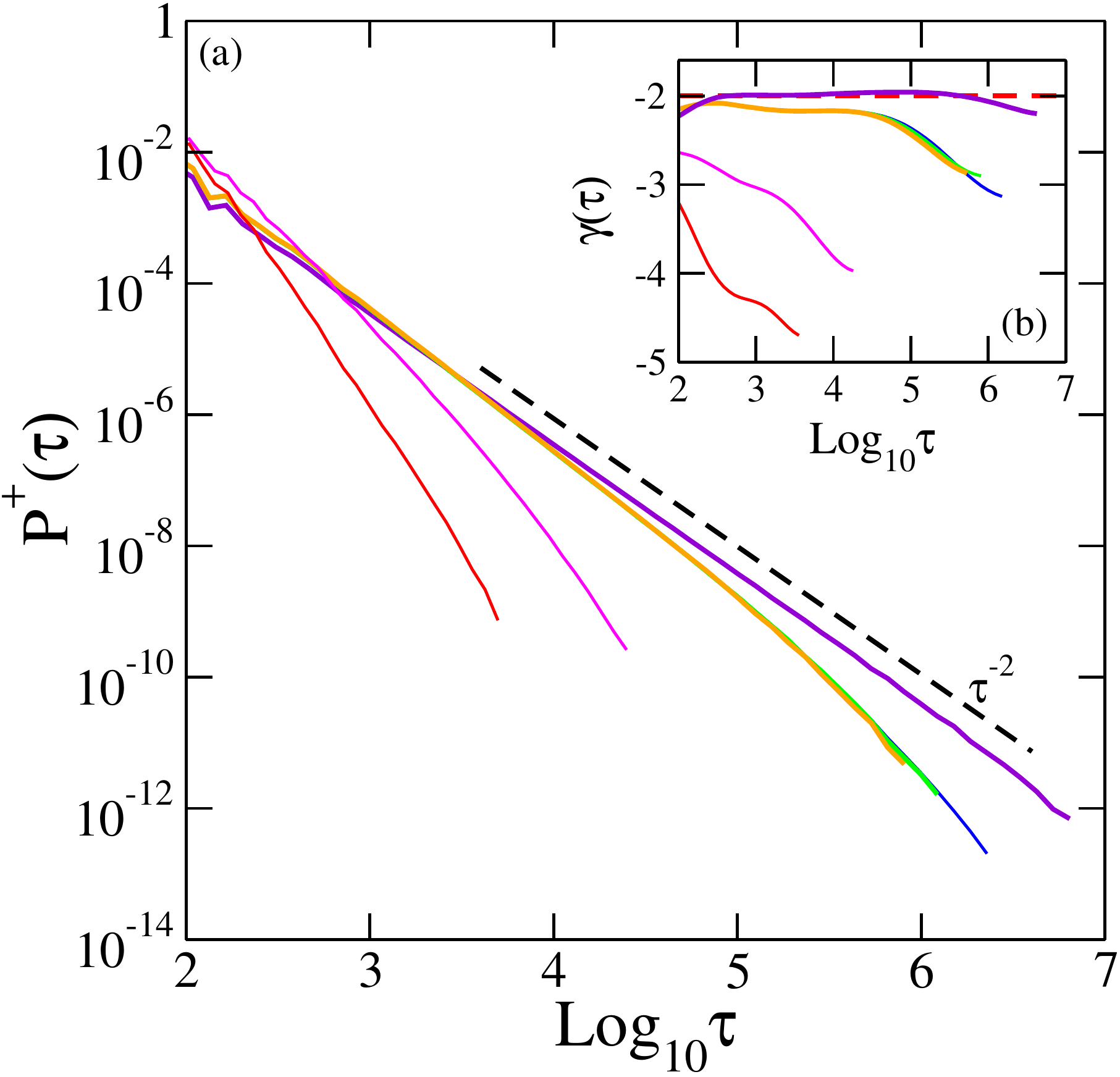}
 \caption{(a) $P^+(\tau)$ obtained for $h = 0.42$ (red), $h = 1.09$ (magenta) and $h = 5.31$ (violet) for $N=2^8$, and $h = 2.5$ for different system sizes $N=2^7$ (orange), $N=2^8$ (green) and $N=2^{10}$ (blue).   
The dashed line guides the eye and indicates the algebraic decay. 
Inset (b): $\gamma(\tau) \equiv d(\log_{10} P^+)/d(\log_{10} \tau)$. 
Here $\varepsilon = 0.05$ and $T=10^{10}$.
}
  \label{fig5}
\end{figure}
\begin{figure}[h]
 \centering
\includegraphics[ width=0.9\columnwidth]{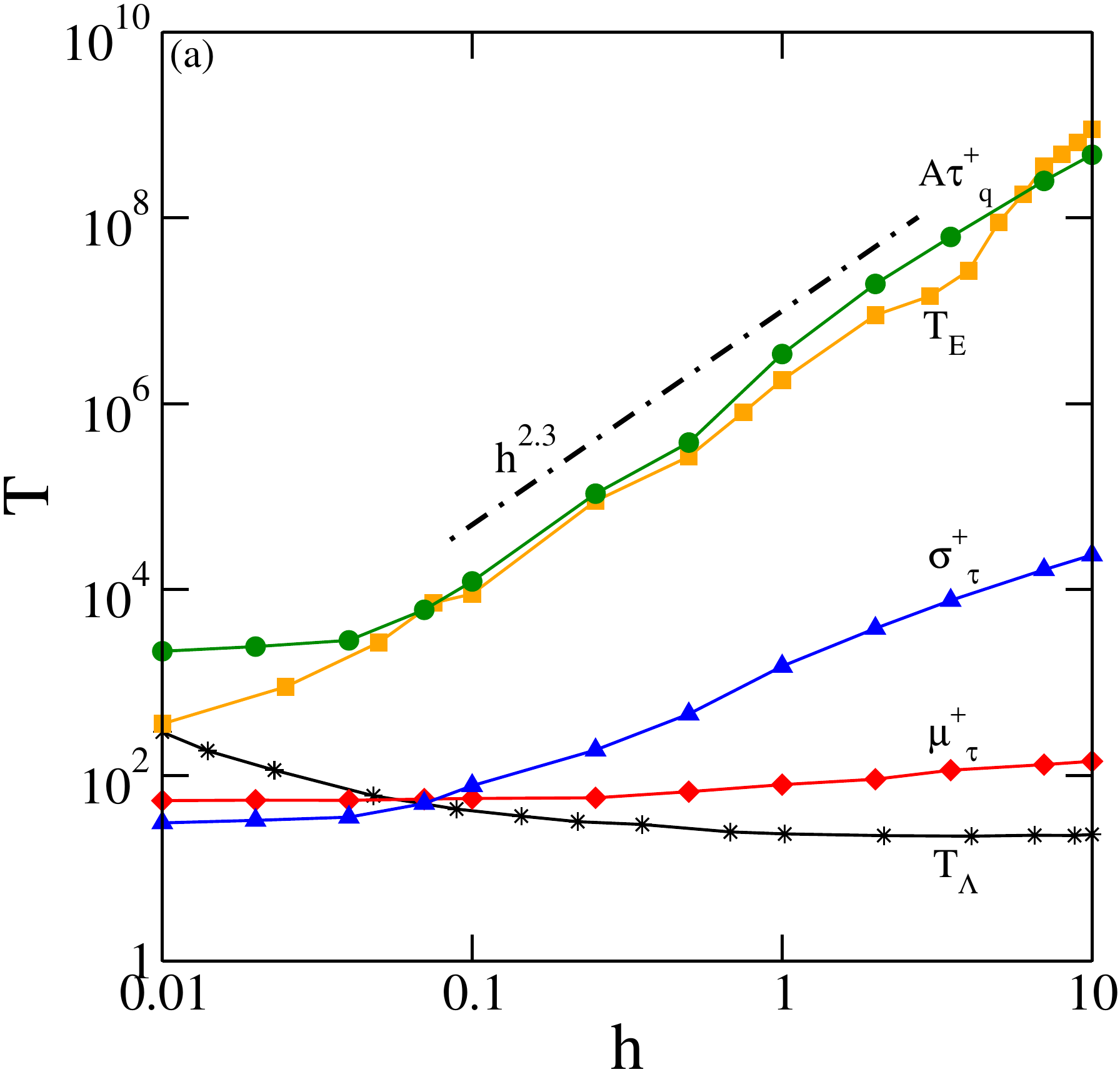}
\includegraphics[ width=0.89\columnwidth]{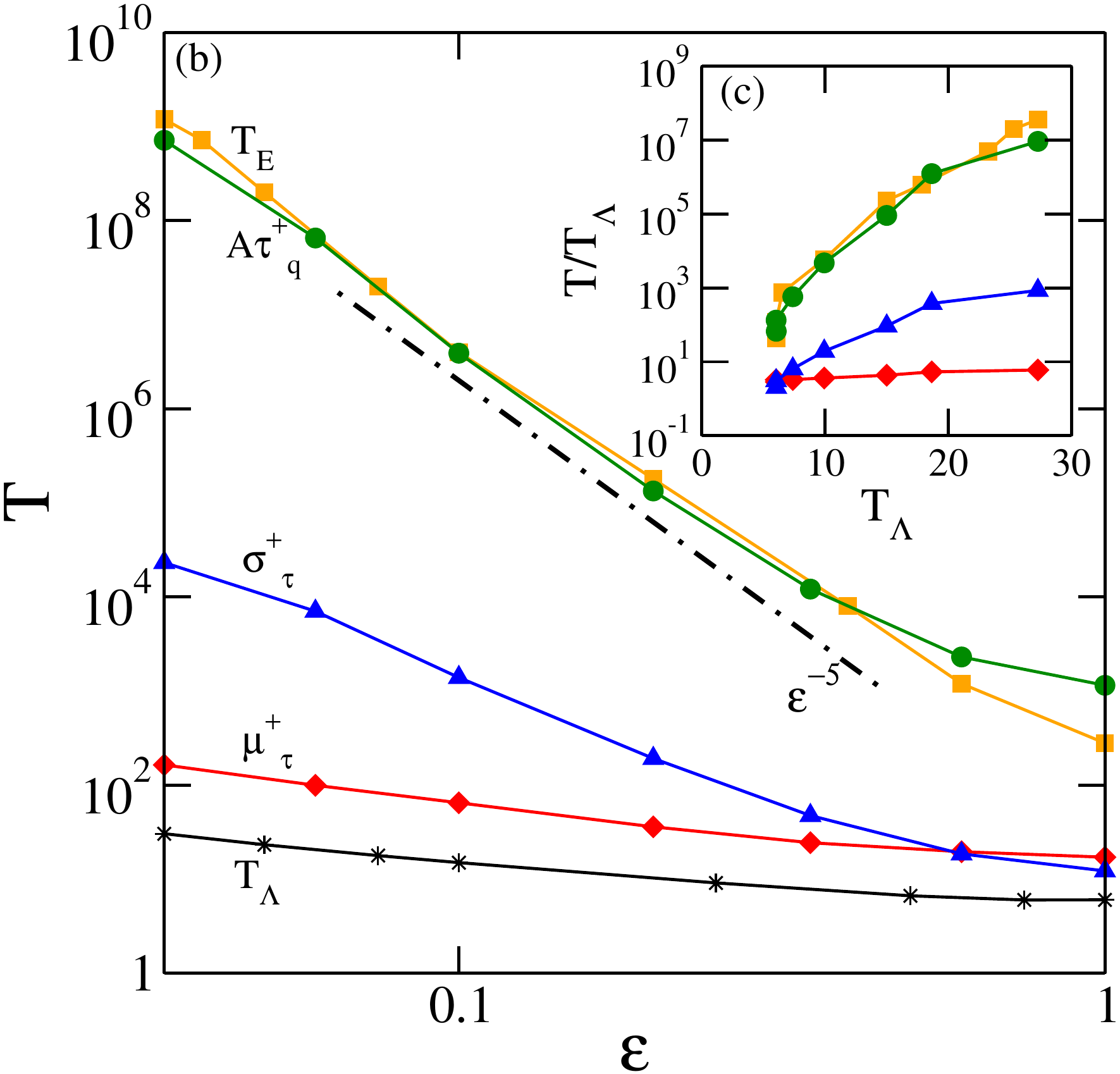}
 \caption{
 (a) Time scales $T_E$ (orange squares), $A\tau_q^+$ (green circles), $\sigma_\tau^+$ (blue triangles), $\mu_\tau^+$ (red diamonds)
and  $T_{\Lambda}$ (black stars) 
vs. energy densities $h$ for $\varepsilon = 0.05$.
 (b) Same as (a) but vs. $\varepsilon$ for $h = 5$ 
 Inset (c) rescaled times ($T_E$, $A\tau_q^+$, $\sigma_\tau^+$, $\mu_\tau^+$) in units of $T_{\Lambda}$. 
The black dashed-dotted lines guide the eye. 
 Here $N=2^{10}$ and $A= 132$. 
}
  \label{fig6}
\end{figure}
\begin{figure}[h]
 \centering
  \includegraphics[ width=0.925\columnwidth]{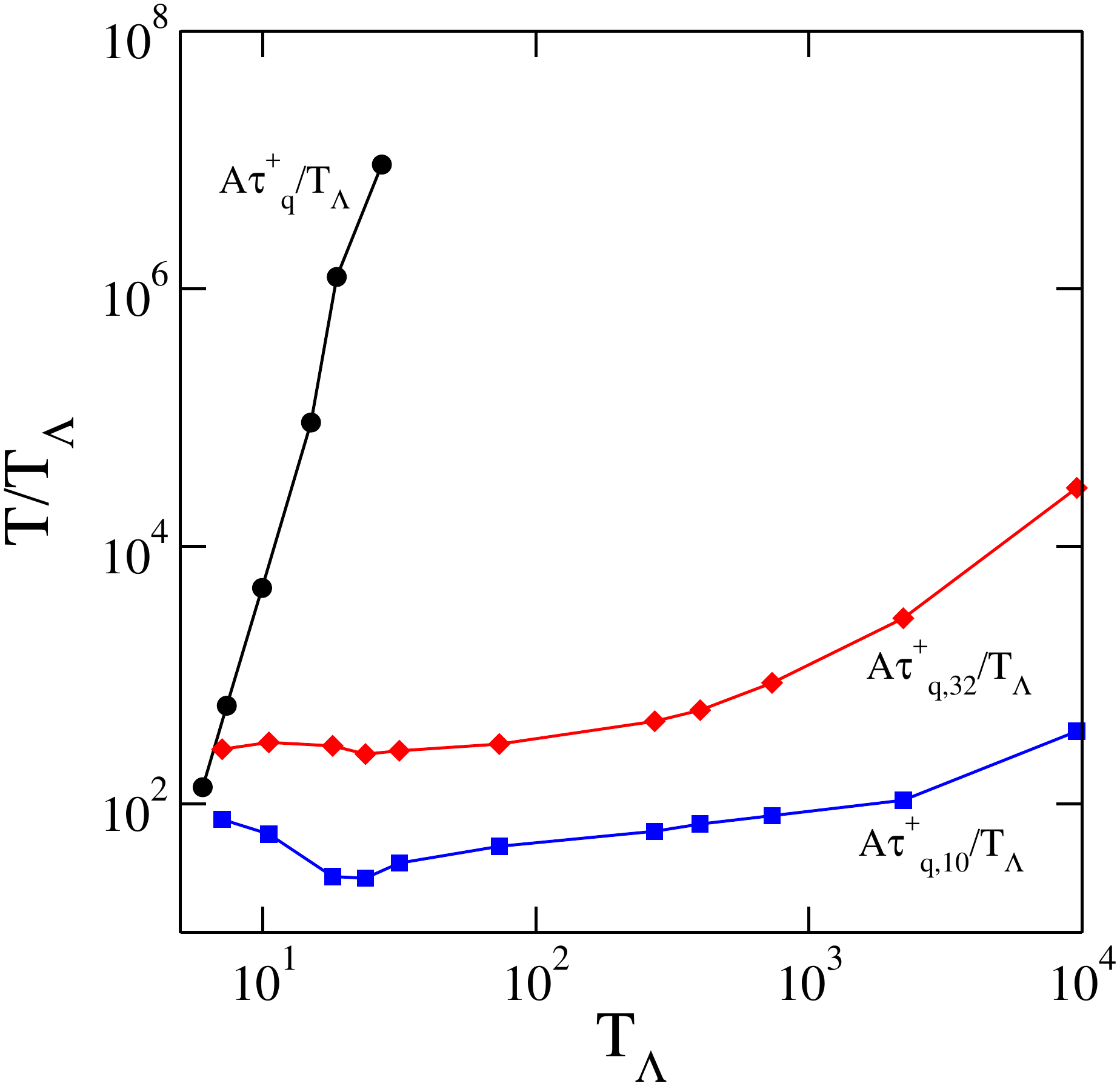}
 \caption{
$A\tau_q^+/T_{\Lambda}$ versus the Lyapunov time $T_{\Lambda}$ for the anti-continuum limit (black circles) with $A= 132$, and the low-energy limit with $k=32$ (red diamonds) and $k=10$ (blue squares), both with $A= 166$.  
}
  \label{fig7}
\end{figure}

In Fig.~\ref{fig4} we show $q(T)$.
Again $q(T)\sim q(0)$ for $T\ll T_E$, 
and $q(T)\sim T_E/T$ for $T\gg T_E$ in accord with Eq.(\ref{eq:qT_TE}). 
The ergodization time $T_E$ is  extracted by rescaling and fitting the curves (see appendix \ref{sec:TE} for details).

We compute the excursion times $\tau_n^+$ of the observables $I_n$, and their distributions $P^+$.
In Fig.~\ref{fig5} we show  $P^+$ for different $h$ with $\varepsilon = 0.05$ with $N=2^8$. We notice that the distributions acquire fat fails, with an intermediate power-law trend $\tau^{-2}$ which extends as $h$ grows.
This is substantiated in the inset, where the local derivative $\gamma(\tau) \equiv d(\log_{10}P^+)/d(\log_{10}\tau)$ is shown.
We also show in Fig.~\ref{fig5} the distributions $P^+$ computed for a given  $h = 2.5$ and $\varepsilon = 0.05$ and different system sizes $N = 2^7,2^8,2^{10}$ (orange, green, blue curves) to confirm the absence of finite size corrections.
We finally compute the average $\mu_\tau^+$  and the standard deviation $\sigma_\tau^+$.

In Fig.~\ref{fig6} we compare all computed time scales for the large energy density regime $h \gg 1,\varepsilon = 0.05$ [plot(a)] and the weak coupling regime  $\varepsilon \ll 1,h=5$ [plot(b)].
We found that $T_E$ grows over several orders of magnitude.
The standard deviation $\sigma_\tau^+$ outgrows
the average $\mu_\tau^+$ as the integrable limit is approached. 
We plot $A\tau_{q}^+$ for $A=132$ (see \cite{ratio_A} for details), 
and observe very good agreement with $T_E$.

Finally we compare $T_E$, $\mu_\tau^+$, $\sigma_\tau^+$ and $A\tau_{q}^+$ with the Lyapunov time $T_\Lambda$ (shown in black stars) as a function of $h$ in  Fig.~\ref{fig6}(a) and $\varepsilon$ in  Fig.~\ref{fig6}(b). 
In contrast to the long range network results where $T_{\Lambda} \approx \sigma_{\tau}^+$ , 
in both short range network cases $T_{\Lambda} \lesssim \mu_{\tau}^+ \ll \sigma_{\tau}^+$.
Consequently,
$T_E\approx 10^9$ (at $h=10$ for given $\varepsilon=0.05$ in Fig.~\ref{fig6}(a), and at $\varepsilon = 0.1$ for $h=5$ in plot(b)), $T_\Lambda \approx 10$,
leaving a gap of eight orders of magnitude in time to be understood.
In Fig.~\ref{fig6}(c), we confirm the above statements by showing $T_E$, $\mu_{\tau,k}^+$, $\sigma_{\tau,k}^+$ and $A\tau_{q,k}^+$ in units of $T_{\Lambda}$, as function of $T_{\Lambda}$.
The observed widening temporal gap between $T_{\Lambda}$ and $T_E$ is similar to the short range network studies of a 
classical chain of Josephson junctions in Ref. \cite{Mithun2018JJ} 
and signals the emergence of a {\it Dynamical Glass} \cite{Mithun2018JJ}.

\section{Conclusion}

Our studies of the microcanonical dynamics of Klein-Gordon chains with up to 1024 degrees of freedom show that the distributions of finite time averages  $\rho(\overline{J}_n;T)$ of integrable limit actions tend towards delta functions in the large $T$ limit.
Consequently the extracted ergodization times $T_E$ increase upon approaching the integrable limits but retain finite values at finite distance from the limits.
We also computed the statistics of fluctuation times of $J_n(t)$. We found that both their average $\mu_{\tau,n}^+$ and standard deviation $\sigma_{\tau,n}^+$
diverge in the same integrable limits, as well as their dimensionless ratio $\sigma_{\tau,n}^+ / \mu_{\tau,n}^+$ which indicates the emergence of fat tails.
Assuming the statistical independence of fluctuation events, it follows that $(\sigma_{\tau,n}^+)^2 / \mu_{\tau,n}^+ \sim T_E$ which was confirmed in all studied cases.
Similar observations were obtained for classical chains of Josephson junctions \cite{Mithun2018JJ}, raising the interesting question as to how general our findings
are.

We studied two different types of integrable limits, characterized by long range respectively short range networks between the actions $J_n$ spanned by the
non-integrable perturbation. 
While the above findings appear to be generic for both cases, the comparison of the Lyapunov time $T_{\Lambda}$ with the
ergodization time scales shows remarkable differences for the two types of networks. For long range networks, we find that $T_{\Lambda} \approx \sigma_{\tau,n}^+$
and consequently $T_E \sim T_{\Lambda}^2$. On the contrary, for short range networks $T_{\Lambda} \approx \mu_{\tau,n}^+$ (see e.g. Fig.~\ref{fig7}).
The latter observation is again in line with similar results obtained for classical chains of Josephson junctions \cite{Mithun2018JJ},  indicating the emergence of a {\it Dynamical Glass}. 

Both types of networks are characterized by a finite coordination number $L$ introduced in section \ref{section2}.
In the case of the KG model, this coordination number amounts to $L=2$ for the SRN and $L=4$ for the LRN cases. 
The Lyapunov time corresponds to the time scale on which the dynamics of resonant interacting multiplets of actions becomes chaotic, in accordance with
Chirikov overlap criterion studies, see e.g. \cite{Livi1987Chaotic,ESCANDE1985stochasticity,Escande1994self_consistent,Lvov2018double,Pistone2018thermalization}). 
Close to the corresponding integrable limit, the probability and corresponding density of resonant multiplets will diminish.
In the case of a LRN, chaotic dynamics of any group of $L$ resonant actions will still couple into the whole network. In contrast, in the case of SRN, the 
chaotic dynamics of a group of $L$ resonant actions will couple only to its nearest neighbors, leaving the dynamics of the majority of all actions 
almost regular and unchanged. We conjecture that the rapidly widening gap between the ergodization time $T_E$ and the Lyapunov time $T_{\Lambda}$ for SRN
is due to slow processes of diffusion, or meandering, or percolation, of resonance through the action network, as observed also for
classical chains of Josephson junctions \cite{Mithun2018JJ}. 

Many open questions remain. 
Among the most pressing ones are the pushing the limits of the calculation to $N \rightarrow \infty$ of the SRN case where the finite size effect on the $q(T)$ disappears. On the other hand, the considered large lattice lattice for this case, $N=1024$, hints that $T_E$ remains to be same even in the limit $N \rightarrow \infty$.
Other challenges concern collecting evidence
that the above observed scenarios of many-body Hamiltonian dynamics approaching integrable limits is generic, and the impact of quantization on the 
slow ergodization dynamics.

\section*{Acknowledgments}

The authors acknowledge financial support from IBS (Project Code No. IBS-R024-D1). We thank I. Vakulchyk, A. Andreanov, M. Fistul and Ch. Skokos for useful discussions.

\appendix

\section{Numerical Integration}\label{app:numerical}

Our simulations were performed on the IBS-PCS cluster, which uses Intel E5-2680v3 processors.
The time integrations were performed using a second order symplectic integrator $SABA_2 C$  (see \cite{Laskar2001high} for a general discussion on symplectic methods; see \cite{Skokos2009Delocalization} for the explicit application of the integrator  $SABA_2 C$ to the KG chain). 
The $SABA_2$ scheme consists in separating the Hamiltonian $H=A+B$, and approximate the resolvent $e^{\Delta t H}$ according to 
\begin{equation}
\begin{split}
SABA_2 = e^{c_1\Delta t L_A}e^{d_1\Delta t L_B}e^{c_2\Delta t L_A}  e^{d_1\Delta t L_B} e^{c_1\Delta t L_A} 
\end{split}
\label{eq:saba2}
\end{equation}
where 
$c_1=\frac{1}{2} ( 1 - \frac{1}{\sqrt{3}} )$, 
$c_2= \frac{1}{\sqrt{3}}  $, 
$d_1=\frac{1}{2}$, 
and $\Delta t$ the time-step. 
We split the Hamiltonian H the KG system in Eq.(\ref{eq:KG_equation2}) as
  \begin{equation}
 A = \sum_{n=1}^{N}\frac{p_n^2}{2}\ ,\quad B =  \sum_{n=1}^{N} \left[ \frac{q_n^2}{2}  +  \frac{q_n^4 }{4}  +  \frac{\varepsilon}{2}(q_{n+1} - q_n)^2 \right] \;.
\label{eq:a1}
\end{equation}
The resolvent operators $e^{\Delta t L_A}$ and $e^{\Delta t L_B}$ of the Hamiltonian $A$ and $B$ propagate the set of coordinates  ($q_n,p_n$)  at the time $t$ to the final values ($q'_n , p_n'$) at the time $t+\Delta t$. These operators  respectively read 
  \begin{eqnarray}
 &   e^{\Delta t L_A}: \left\{
                \begin{array}{ll}
                  q_n' =q_n+ p_n \Delta t\\
                   p_n' =p_n
                \end{array}
                    \label{eq:LA}
              \right. \\
   &               e^{\Delta t L_B}:\left\{
          \begin{array}{ll}
              q_n' =q_n\\
                p_n' = p_n +   \big\{  - q_n(1 + q_n^2) \\
                & \hspace{-35mm}  +   \varepsilon (q_{n+1} + q_{n-1} -2 q_n )  \big\} \Delta t \\
          \end{array}
          \right.
              \label{eq:LB}
  \end{eqnarray}
Following \cite{Skokos2009Delocalization}, we improve the accuracy of the $SABA_2$ scheme using a corrector $C=\{ \{A,B\} , B\}$. 
\begin{equation}
\begin{split}
\text{SABA}_2 C  =  e^{-\frac{g}{2}\Delta t^3 L_C}  \text{SABA}_2    e^{-\frac{g}{2}\Delta t^3 L_C}
\end{split}
\label{eq:saba2c}
\end{equation}
for $g = (2 - \sqrt{3})/24$. For the KG chain, the corrector $C$ is 
\begin{equation}
\begin{split}
C &=  \sum_{n=1}^{N}\Big[ q_n(1 + q_n^2) +   \varepsilon (2 q_n - q_{n+1} - q_{n-1} )   \Big]^2\ .
\label{eq:corrector}
\end{split}
\end{equation}
The corrector operator $C$ yields the following resolvent operator 
      \begin{widetext}
  \begin{equation}
                  e^{\Delta t L_C}:\left\{
          \begin{array}{ll}
              q_n' =q_n\\
%
  p_n' =  p_n +  2 \Big\{ \big[ - q_n(1 + q_n^2)  +  \varepsilon(  q_{n+1} + q_{n-1} - 2 q_n) \big]\big[ 1 + 3 q_n^2  +   2\varepsilon   \big]  \\
         \quad  + \varepsilon \big[ q_{n-1}(1 + q_{n-1}^2)  -   \varepsilon ( q_{n} + q_{n-2} - 2 q_{n-1}) \big]\\
          \quad + \varepsilon \big[  q_{n+1}(1 + q_{n+1}^2)   -  \varepsilon ( q_{n+2} + q_{n} -  2 q_{n+1})\big]  \Big\} \Delta t  
          \end{array}
          \right.
              \label{eq:LC}
  \end{equation}
 \end{widetext}
Note that for both resolvents  $e^{\Delta t L_B}$ and $e^{\Delta t L_C}$ in Eq.(\ref{eq:LB}) and Eq.(\ref{eq:LC}) the boundary conditions have to be applied: fixed boundary condition for the LRN cases, and periodic boundary conditions for the SRN cases.
This scheme was implemented
using a time-step $\Delta t = 0.1$, keeping the relative energy error $\Delta E = |E(t) - E(0)|/E(0)$ of order $10^{-6}$.

\section{Distribution $P_k$ of the excursion times}\label{app:PDF}

In the numerical calculations of the PDF $P_k$ of excursions out of equilibrium $\tau$, we consider the interval $\mathcal{I}_b = [10^2,10^b]$ to which
the excursion times belong, and we separate this into $M$ bins of logarithmic width \cite{Clauset2009PowerLawDI}.  
Hence, the bins $B_s$ are defined as
\begin{equation}
\mathcal{I}_b = \cup_{s=1}^M \mathcal{B}_s  = \cup_{s=1}^M [10^{2+\kappa (s-1)},10^{2+\kappa s}]
\label{eq:bins}
\end{equation}
with $\kappa = (b-2)/M$. 
From our plots, we exclude all the bins that count fewer
than 100 events, as these should be  considered as statistically not relevant.

\section{Average and standard deviation of $\tau_n^\pm$}\label{sec:tau_pm}

\begin{figure}[h]  
 \centering
\includegraphics[ width=0.9\columnwidth]{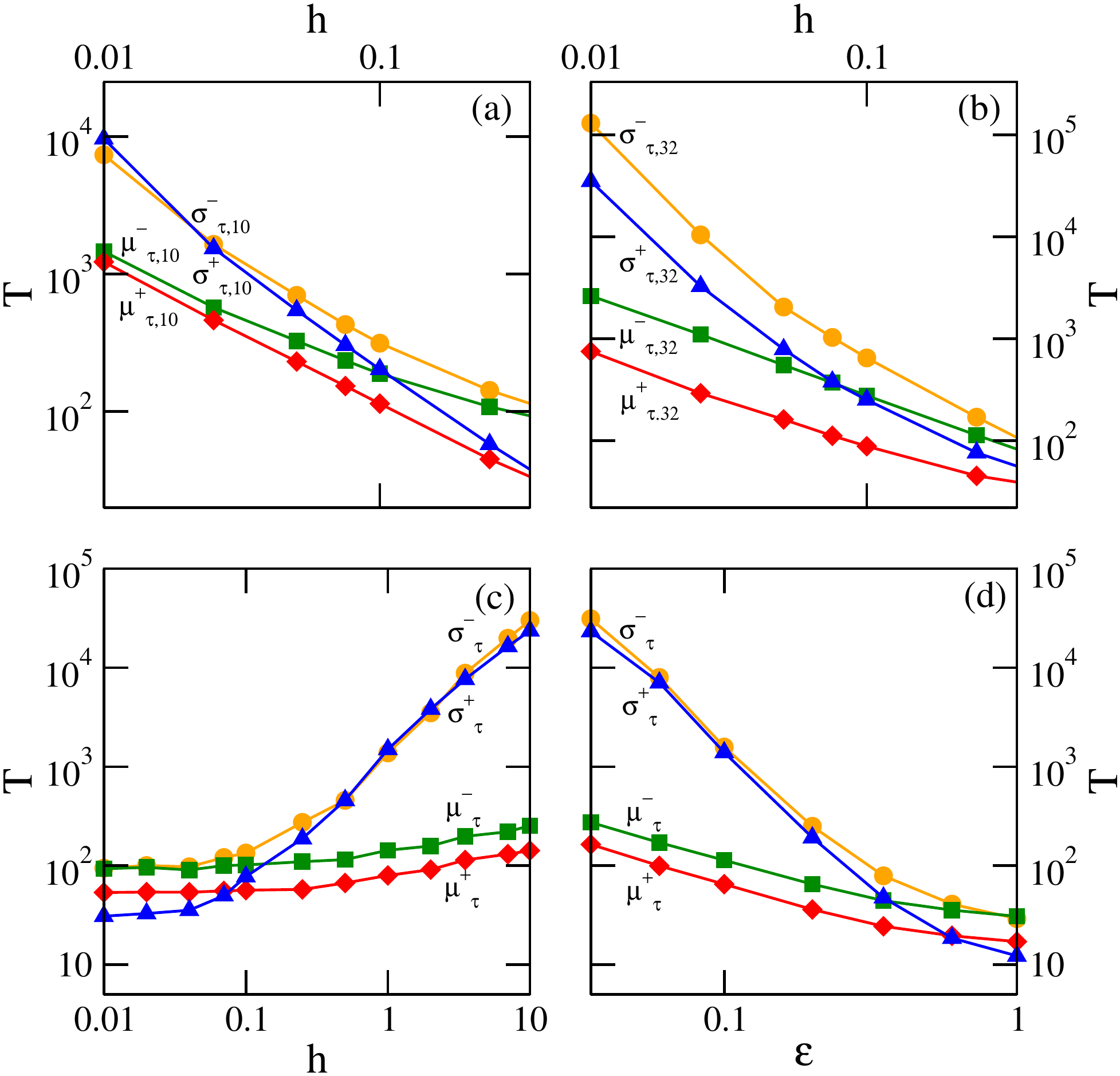}
 \caption{
  (a) $\mu_{\tau,k}^+$ (red diamonds), $\mu_{\tau,k}^-$ (green squares), $\sigma_{\tau,k}^+$ (blue triangles) and $\sigma_{\tau,k}^-$ (orange circles) versus the energy density $h$ for $k=10$ and $\varepsilon = 1$.
 (b) Same as (a) for $k=32$. 
 In both (a) and (b), $M=2^9$. 
 (c) $\mu_\tau^+$ (red diamonds), $\mu_\tau^-$ (green squares), $\sigma_\tau^+$ (blue triangles) and $\sigma_\tau^-$ (orange circles) versus the energy density $h$ for $\varepsilon = 0.05$.
 (d) Same as (c) versus $\varepsilon$ for $h=5$.
}
  \label{fig5_app}
\end{figure}
In Fig.~\ref{fig5_app} we show the averages and the standard deviations of both $\tau_k^+$ and $\tau_k^-$. 
The four plots (a-d) corresponds to the four cases discussed numerically in the main text, namely: 
$\mu_{\tau,k}^\pm$ and $\sigma_{\tau,k}^\pm$ for the low energy regime with $k=10$ (a) and for $k=32$ (b), corresponding to Fig.~\ref{fig3}(a) and Fig.~\ref{fig3}(b) respectively;  
$\mu_\tau^\pm$ and $\sigma_\tau^\pm$ for the large energy limit (c) and anti-continuum limit (d) corresponding to Fig.~\ref{fig6}(a) and Fig.~\ref{fig6}(b) respectively. 
In all these cases, we observe that the average $\mu_\tau^+$ of $\tau^+$ (red diamonds) shows a divergence trend similar to
$\mu_{\tau}^-$ of $\tau^-$ (green squares). Similarly,
the standard deviation $\sigma_\tau^+$ of $\tau^+$ (blue triangles) shows a divergence trend like that of
$\sigma_\tau^-$ of $\tau^-$ (orange circles).
We then focus on the $"+"$ events only $\tau^+$, which we recall are events during which $J_n > \langle J_n\rangle_X$.

\section{Ensemble of initial conditions}\label{app:IC}

Let us assume that there exists only one conserved quantity of the system (the total energy $H$). We define the initial condition at $t=0$ as zeroing the position coordinates $q_n = 0$ and distributing the half squares of kinetic energy $p_n^2 / 2$, according to the following distribution $P_1$ defined for a positive real number $C> 0$
\begin{equation}
P_1(z) =C e^{-C z}\ ,\qquad z\in ]0,\infty[\ .
\label{eq:Max_distro_IC}
\end{equation} 
From this, for a uniform distribution of random numbers $w(r)$ distributed in the range $[0,1]$, one can get  $z =-log(w(r))/C$. This leads to a set of initial momenta coordinates $p_n^{(1)}$ at $t=0$, where the sign is a discrete random variable, $a_n = \pm 1 $ with distribution $P_2(a_n = \pm 1) = 0.5$. Here
\begin{equation}
p_n^{(1)} = a_n \sqrt{2 z} \ . 
\label{eq:Max_distro_IC_3}
\end{equation}
The total energy $E_T$ of the system is
\begin{equation}
E_T = \sum_{n=1}^N \frac{\Big(p_n^{(1)}\Big)^2}{2}  \ .
\label{eq:IC_2}
\end{equation}
We then set a chosen energy density $h$ by the following rescaling 
\begin{equation}
p_n^{(2)} =   \sqrt{\frac{h N}{E_T}}  p_n^{(1)}\ .
\label{eq:IC_1}
\end{equation}
The resulting momentum coordinates $p_n^{(2)}$ with the position coordinates $q_n = 0$ fixed at zero are evolved in time for $T_{IC} = 10^5$ using the $SABA_2C$ integrator with time-step $\tau = 10^{-2}$, which keeps the relative energy error at $\Delta E \sim 10^{-9}$.  
The result of this time evolution is then taken as an initial condition of our simulation. 
Then, $M$ draws of the distributions $P_1$ and $P_2$ yield an ensemble of $M$ initial conditions.

\section{Asymptotic behavior of $q(T)$}\label{sec:q_1/T}

\begin{figure}[h]
 \centering
 \includegraphics[ width=\columnwidth]{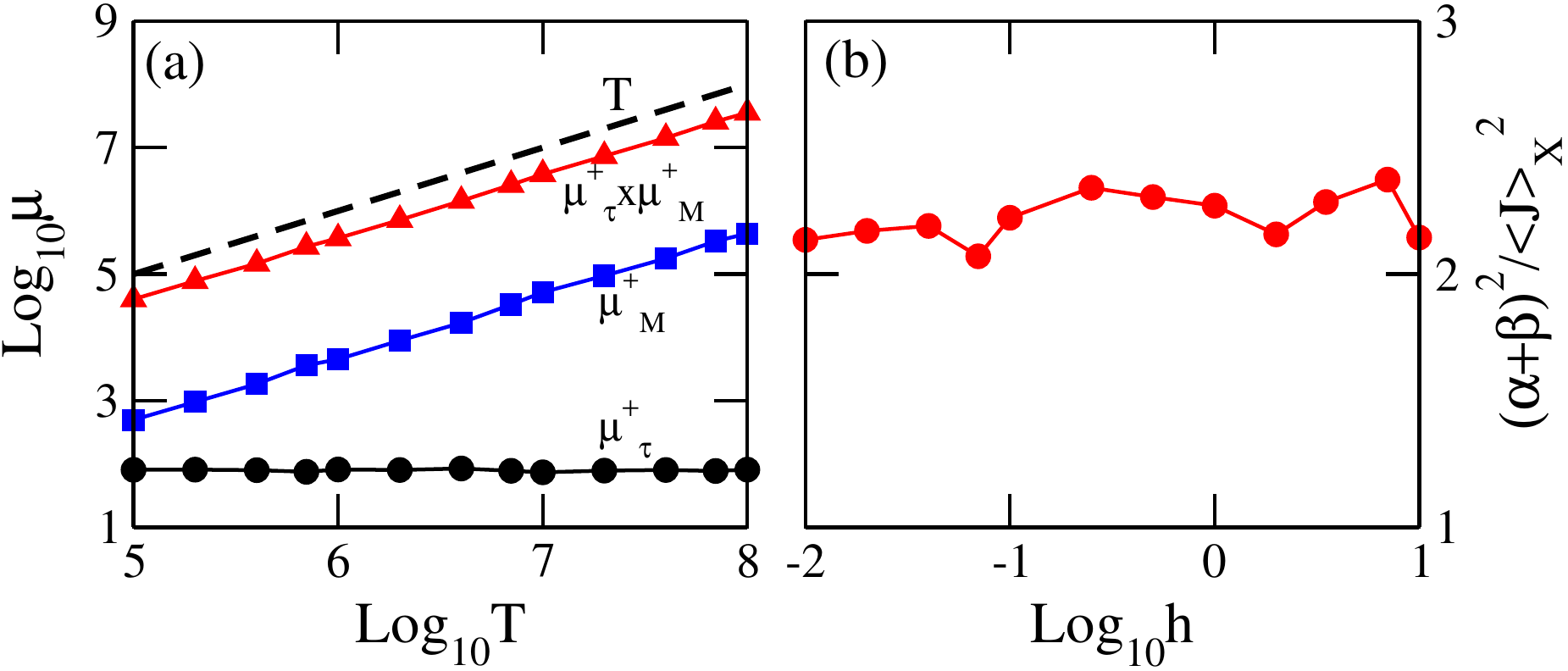}
 \caption{
(a) $\mu_\tau^+$ (black circles), $\mu_{M}^+$ (blue squares) and their product $\mu_\tau^+\times \mu_{M}^+$ (red), versus the integration time $T$ obtained for $h=5$ and $\varepsilon = 0.05$.
The dashed line guides the eye and indicates the linear growth $T$. 
(b): $(\alpha + \beta)^2 / \langle J\rangle_X^2$ versus the energy density $h$ for $\varepsilon = 0.05$.
Here $N=2^{10}$. 
}
  \label{fig2_app}
\end{figure}

In Eq.(\ref{eq:qT_asymp})
we indicated that the ergodization time $T_E$ of an observable $J_n$ is proportional to the 
ratio between the variance $\big(\sigma_{\tau,n}^+ \big)^2$ and the average $\mu_{\tau,n}^+$ 
of its excursions out of equilibrium 
\begin{equation}
T_E \sim 
\tau_{q,n}^+ \equiv 
\frac{ \big( \sigma_{\tau,n}^+ \big)^2}{\mu_{\tau,n}^+}\ .
\label{eq:qT_asymp_app}
\end{equation}
We here derive
this relation, which is based on the approximation of the time evolution of $J_n$ with telegraphic random signal \cite{richard1972feynman,Chakravarty1985photoinduced,kac1963van}. 
Away from the integrable limit, an action $J_n$ becomes a time-dependent variable $J_n(t) = \langle J_n\rangle_X + \phi_n(t)$ where $\phi_n$
is a continuous function fluctuating around zero. 
At the piercing times  $t_n^i$ of the action $J_n$, it follows that  $\phi_n(t_n^i) = 0$. 
Then, the time average of $J_n$,
here indicated as $B_n(T)$ is 
\begin{equation}
\begin{split}
B_n(T) &= \frac{1}{T}\int_0^TJ_n(t) dt = \langle J_n\rangle_X + \frac{1}{T}\int_0^T\phi_n(t) dt \\
&\equiv \langle J_n\rangle_X + D_n(T)\ .
\end{split}
\label{eq:Jk_Tave_app}
\end{equation}
The interval $[0,T]$ consists in $M_n^+$ events $\tau_n^+$ and $M_n^-$ of $\tau_n^-$, plus uncompleted
initial and final
events of duration $\mathcal{I}_n$ and $\mathcal{E}_n$ respectively \cite{Wang2018renewal}.
Let us observe that both $\mathcal{I}_n$ and $\mathcal{E}_n$ are not distributed according to the distribution $P_n^\pm$ of excursion times $\tau_n^\pm$. 
Nevertheless, for large values of $M_n^+$ and $M_n^-$ we 
neglect their contribution. Hence, for all $k$ it holds that 
\begin{equation}
\begin{split}
T &=\sum_{i=1}^{M_n^+ } \tau_{n}^+(i) + \sum_{i=1}^{M_n^-} \tau_{n}^-(i) \ .
\end{split}
\label{eq:T_split}
\end{equation} 
The numbers $M_n^\pm$ of events  $\tau_n^\pm$ are distributed according to the distributions $\rho_{M_n}^\pm$ and have an average $\mu_{M_n}^\pm$. 
In Fig.~\ref{fig2_app}(a), we show that the average $\mu_{M_n}^+$ of the $\tau_n^+$ scales typically as $\mu_{M_n}^+ \sim T / \mu_{\tau_n}^+$ as $T\rightarrow \infty$, where $\mu_{\tau_n}^+$ is the average of $\tau_n^+$ computed for the short range case of the KG chain using the observables $J_n$ in Eq.(\ref{eq:KG_observables}). 
From Eq.(\ref{eq:T_split}), we can write the integral over the interval $[0,T]$ in Eq.(\ref{eq:Jk_Tave_app}) as a sum of integrals over the excursion times
\begin{equation}
\begin{split}
\int_0^T\phi_n(t) dt &= 
 \sum_{i=1}^{M_n^+ }  \int_{t_n^i}^{t_n^{i+1}}\varphi_n^+(t) dt -  \sum_{i=1}^{M_n^- }  \int_{t_n^{i+1}}^{t_n^{i+2}}\varphi_n^-(t)dt \\
&\equiv  \alpha_n  \sum_{i=1}^{M_n^+ } \tau_{n}^+(i) - \beta_n \sum_{i=1}^{M_k^- } \tau_{n}^-(i)  
\end{split}
\label{eq:phi_k_int_app}
\end{equation}
where $\alpha_n$ and $\beta_n$ are defined as
\begin{equation}
\begin{split}
 \alpha_n &=\frac{ \sum_{i=1}^{M_n^+ }  \int_{t_n^i}^{t_n^{i+1}} \varphi_n^+(t) dt }{\sum_{i=1}^{M_n^+ } \tau_{n}^+(i)} \ , 
 \quad
 \beta_k  =\frac{ \sum_{i=1}^{M_n^- }  \int_{t_n^i}^{t_n^{i+1}} \varphi_n^-(t) dt }{\sum_{i=1}^{M_n^-} \tau_{n_n}^-(i)} \ .
\end{split}
\label{eq:alpha_beta_n}
\end{equation}
These coefficients $\alpha_n$ and $\beta_{n}$ are distributed by the distributions $\rho_{\alpha_n}$ and $\rho_{\beta_n}$ respectively. Let us here define their averages $\alpha$ and $\beta$.
We then approximate Eq.(\ref{eq:phi_k_int_app}) by
the telegraphic noise signal
\begin{equation}
\begin{split}
\int_0^T\phi_n(t) dt &\approx \alpha  \sum_{i=1}^{M_n^+ } \tau_{n}^+(i) - \beta \sum_{i=1}^{M_n^- } \tau_{n_n}^-(i) \\
 & \equiv \alpha S_n^+ - \beta S_n^-\ .
\end{split}
\label{eq:phi_k_int_app_2}
\end{equation}
Let us now consider the limit of $q(T)$ for $T\rightarrow\infty$. 
Due to the continuity of $\phi_n$ and the finiteness of all moments of the excursion times $\tau_n^\pm$, the term $D_n(T)$ in Eq.(\ref{eq:Jk_Tave_app}) converges to zero as $T\rightarrow \infty$. 
Then, it follows that $ \lim_{T\rightarrow \infty} \mu_{J_n}^2(T) = \langle J_n\rangle_X^2$. 
Hence, supposing $\langle J_n\rangle_X\neq 0$, the limit of the index $q$ is 
\begin{equation}
\begin{split}
\lim_{T\rightarrow \infty} q(T) &=  \lim_{T\rightarrow \infty} \frac{1}{\mu_{J_n}^2(T)} \cdot \lim_{T\rightarrow \infty}  \sigma_{J_n}^2(T) \\
&= \frac{1}{\langle J_n\rangle_X^2} \cdot \lim_{T\rightarrow \infty}  \sigma_{J_n}^2(T) \ .
\end{split}
\label{eq:q_lim1}
\end{equation}
Recalling the following properties of the variance for a constant $A$
\begin{equation}
\sigma_{ A J_n}^2 = A^2 \sigma_{J_n}^2\ , \qquad 
\sigma_{A+ J_n}^2  = \sigma_{J_n}^2
\label{eq:var_properties}
\end{equation}
then from Eq.(\ref{eq:Jk_Tave_app}) it follows that
\begin{equation}
\sigma_{ B_n(T)}^2 = \sigma_{ D_n(T)}^2\ .
\label{eq:var_Bk1}
\end{equation}
We can restrict to
the $\tau_n^+$ events only in Eq.(\ref{eq:phi_k_int_app_2}) 
by adding and subtracting $\beta S_n^+$. It follows that
\begin{equation}
\begin{split}
D_n(T)&=  \frac{1}{T}\int_0^T\phi_n(t) dt =  \frac{1}{T}\left[  \alpha S_n^+ - \beta S_n^- \left( \pm \beta S_n^+ \right) \right] \\
&= \frac{1}{T}\left[  \left( \alpha + \beta\right)  S_n^+  - \beta  \left(  S_n^- +  S_n^+ \right) \right]\\
&= \frac{1}{T}\left[   \left( \alpha + \beta\right)  S_n^+  - \beta T\right] 
=  \frac{ \alpha + \beta }{T}  S_n^+  - \beta\ .
\end{split}
\label{eq:phi_k_int_prop}
\end{equation}
By Eq.(\ref{eq:var_properties}), the variance of $B_n$ is
\begin{equation}
\sigma_{ B_n(T)}^2 =  \frac{ (\alpha + \beta)^2 }{T^2}  \sigma_{ S_n^+}^2 \ .
\label{eq:var_Bk1}
\end{equation}
The excursion times $\tau_n^+$ are identically distributed variables.  
Assuming these to be independent events, the variance $ \sigma_{ S_n^+}^2 $ of the sum of the head events $S_n^+$ in Eq.(\ref{eq:phi_k_int_prop}) is the product of the variance $(\sigma_{ \tau,n}^+)^2$ of the head events multiplied by the number of events $M_n^+$
\begin{equation}
S_n^+\equiv \sum_{i=1}^{M_n^+}  \tau_n^+(i) 
\quad\Rightarrow\quad 
 \sigma_{ S_n^+}^2  = M_n^+  (\sigma_{ \tau,n}^+)^2\ .
\label{eq:var_tau}
\end{equation}
For $T\gg \mu_{\tau,n}^+$, we expect that
$\mu_{M_n}^+ \sim T / \mu_{\tau,n}^+$.
In Fig.~\ref{fig2_app}(a) we report the averages $\mu_{M}^+$ and $\mu_\tau^+$ of the number $M_n^+$ and the duration $\tau_n^+(i)$ of the events detected by all the observables $J_n$ in Eq.(\ref{eq:KG_observables}) for $h=5$ and $\varepsilon = 0.05$. 
This plot confirms the above expectation.
\\

\noindent
In Eq.(\ref{eq:var_tau}) we approximate the variable $M_n^+$ by
its average $\mu_{M_n}^+$, which leads
to
\begin{equation}
\sigma_{ B_n(T)}^2 \sim   (\alpha + \beta)^2 \frac{ (\sigma_{ \tau,n}^+)^2 }{\mu_{\tau,n}^+}     \frac{1 }{T}  \ .
\label{eq:var_Bk2}
\end{equation}
Ultimately, this results in the formula
\begin{equation}
q(T) \sim  \frac{ (\alpha + \beta)^2 }{ \langle J_n\rangle_X^2 }  \frac{ (\sigma_{ \tau,n}^+)^2 }{\mu_{\tau,n}^+} \frac{ 1 }{ T} \ .
\label{eq:q_asympt_app}
\end{equation}
In Fig.~\ref{fig2_app}(b) we show the energy  density $h$ dependence of the ratio $(\alpha + \beta)^2 /  \langle J_n\rangle_X^2$ introduced in Eq.(\ref{eq:q_asympt_app}) computed for the short range case of the KG chain using the observables $J_n$ in Eq.(\ref{eq:KG_observables}) for $\varepsilon = 0.05$.
%
Over three orders of magnitude, the ratio $(\alpha + \beta)^2 /  \langle J_n\rangle_X^2$ fluctuates between $2$ and $2.5$. Hence this ratio in Eq.(\ref{eq:q_asympt_app}) does not contribute to the asymptotic trend of the fluctuation parameter $q(T)$ as the system approaches an integrable limit. 
It therefore
follows that the time-scale $\sigma_{ \tau,n}^+)^2 / \mu_{\tau,n}^+$ is proportional to the ergodization time $T_E$ defined in Eq.(\ref{eq:qT_TE}), as stated in Eq.(\ref{eq:qT_asymp}).

\section{Largest Lyapunov exponent $\Lambda$} \label{app:Lyapunov}

We compute the largest Lyapunov exponent  $\Lambda$ by
considering 
a small amplitude deviation vector $ w(t) =  (\delta q(t) , \delta p(t))$ of a trajectory. 
We then numerically solve the variational equations  
\begin{equation}
\dot{ w}(t) =\big[  J_{2N} \cdot D_H^2 ( x(t) ) \big] \cdot w(t)
\label{eq:var_eq}
\end{equation}
associated with a Hamiltonian $H$ using the tangent method \cite{Skokos2010numerical,Gerlach2011comparing,gerlach2012efficient}. 
where $D_H^2$ is the Hessian matrix and $J_{2N}$ the symplectic matrix. 
Solving Eq.(\ref{eq:var_eq}) using the $SABA_2C$ integration scheme presented in Appendix \ref{app:numerical} yield extended resolvent opertators $e^{\Delta t L_A}$, $e^{\Delta t L_B}$ and $e^{\Delta t L_C}$ in Eq.(\ref{eq:LA},\ref{eq:LB},\ref{eq:LC}), where 
 ($\delta q_n, \delta p_n$)  at the time $t$ are simultaneously integrated to  ($\delta q'_n , \delta p_n'$) at the time $t+\Delta t$. 
The additional equations for $e^{\Delta t L_A}$ are
\begin{equation}
\begin{split}
e^{ \Delta t L_{AV}}: \left\{
\begin{array}{rl}
  \delta q_n' &=   \delta q_n + \delta p_n \Delta t   \\
    \delta p_n' &= \delta  p_n
\end{array} \right. 
\end{split}
\label{eq:LAV}
\end{equation}
while for $e^{\Delta t L_B}$ are
\begin{equation}
\begin{split}
e^{ \Delta t L_{BV}}: \left\{
\begin{array}{rl}
  \delta q_n' &=   \delta q_n + \delta p_n \Delta t   \\
    \delta p_n' &=   \delta  p_n + 
     \big\{  \varepsilon (\delta  q_{n+1} +  \delta  q_{n-1}) \\
     &\qquad  - [ 1 + 3q_n^2 + 2\varepsilon  ]\delta  q_n   \big\}\Delta t 
\end{array} \right. 
\end{split}
\label{eq:LBV}
\end{equation}
For the correction term, we get
\begin{equation}
\begin{split}
e^{ \Delta t L_{CV}}: \left\{
\begin{array}{rl}
    \delta q_n' &=  \delta  q_n   \\
    \delta p_n' &= \delta p_n + \big\{ \gamma_n \delta  q_{n}  +  \gamma_{n+1} \delta  q_{n+1} 
     \gamma_{n+2} \delta  q_{n+2} \\&\quad +  \gamma_{n-1} \delta  q_{n-1} +  \gamma_{n-2} \delta  q_{n-2}   \big\}\Delta t 
\end{array} \right.
\end{split}
\label{eq:LCV}
\end{equation}
where 
\begin{equation}
\begin{split}
\gamma_{n}&= -2 \Big\{ \big[ 1 + 3 q_n^2 +   2\varepsilon  \big]^2 \\&\quad  + 6 q_n \big[ q_n(1 + q_n^2) \\ &\quad +   \varepsilon(2 q_n - q_{n+1} - q_{n-1} ) \big] +2\varepsilon^2 \Big\} \\
\gamma_{n+1}&= 2\varepsilon \big[ 2 + 4\varepsilon + 3 q_n^2 + 3q_{n+1}^2  \big]    \\
\gamma_{n-1}&=   2\varepsilon \big[ 2 + 4\varepsilon + 3 q_n^2 + 3q_{n-1}^2  \big]  \\
\gamma_{n+2}&=  -2  \varepsilon^2   \\
\gamma_{n-2}&=  -2  \varepsilon^2 
\end{split}
\label{eq:LCV_coeff}
\end{equation}
As mentioned above, in both resolvents  $e^{\Delta t L_{BV}}$ and $e^{\Delta t L_{CV}}$ in Eq.(\ref{eq:LBV}) and Eq.(\ref{eq:LCV}) the boundary conditions have to be applied: fixed boundary condition for the LRN cases, and periodic boundary conditions for the SRN cases.
The largest Lyapunov exponent $\Lambda$ is computed by considering the limit
\begin{equation}
\Lambda=
\lim_{t\rightarrow \infty}  \frac{1}{t}\ln \frac{\| w(t)\|}{\| w(0)\|}
\label{eq:Lyap_1}
\end{equation} 
where $\| \cdot \|$ is a norm of the vector $w$. 
An extended presentation of this method applied to the KG chain can be found in \cite{Many2017Computational}.

\section{Measurement of the coefficient $T_E$}\label{sec:TE}

In both Fig.~\ref{fig3} and Fig.~\ref{fig6} we have shown the behavior of the ergodization time $T_E$  as the system approaches the integrable limit.

\begin{figure}[h]  
 \centering
\includegraphics[ width=0.9\columnwidth]{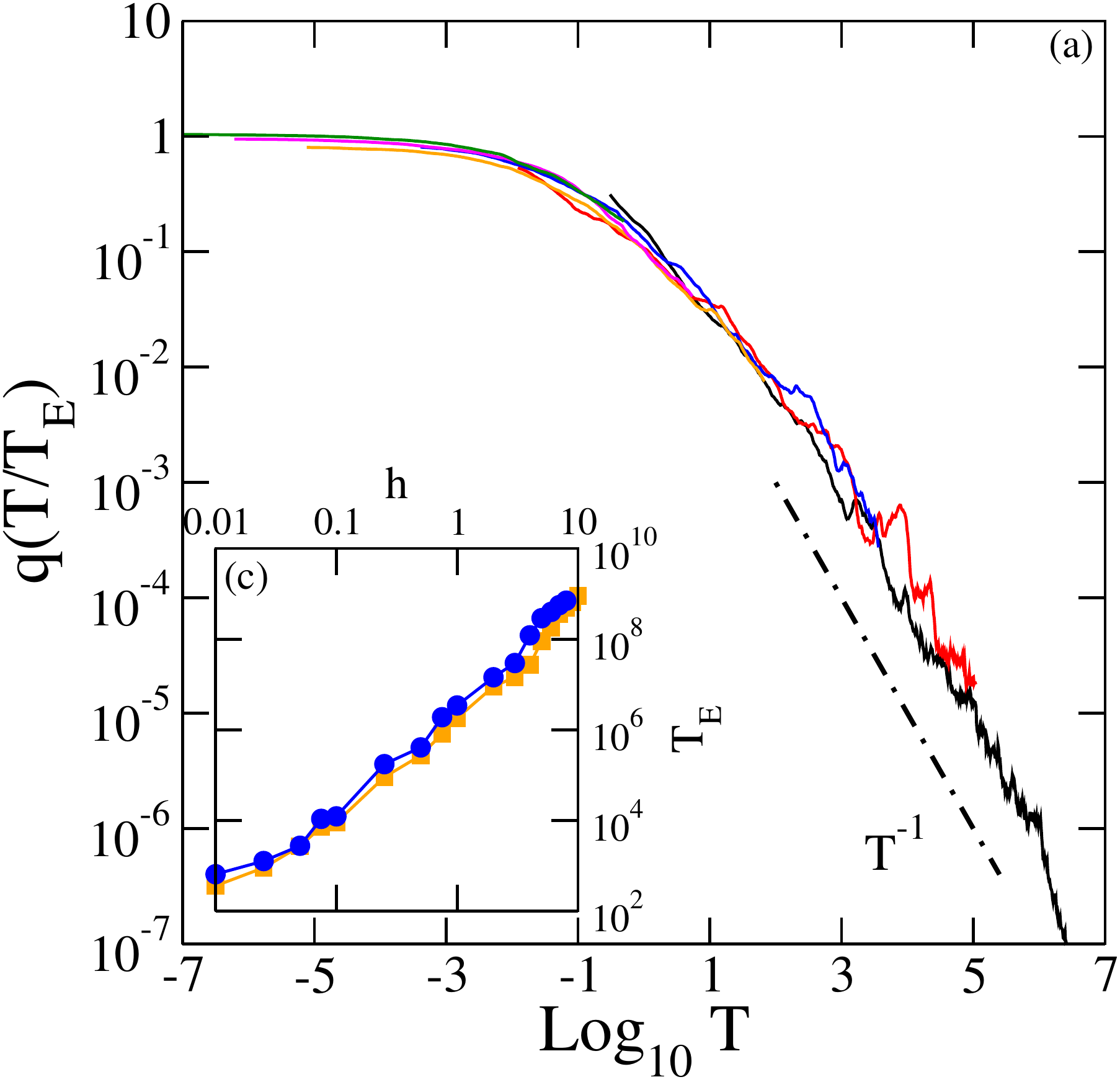}
\vspace{-05mm}
\includegraphics[ width=0.9\columnwidth]{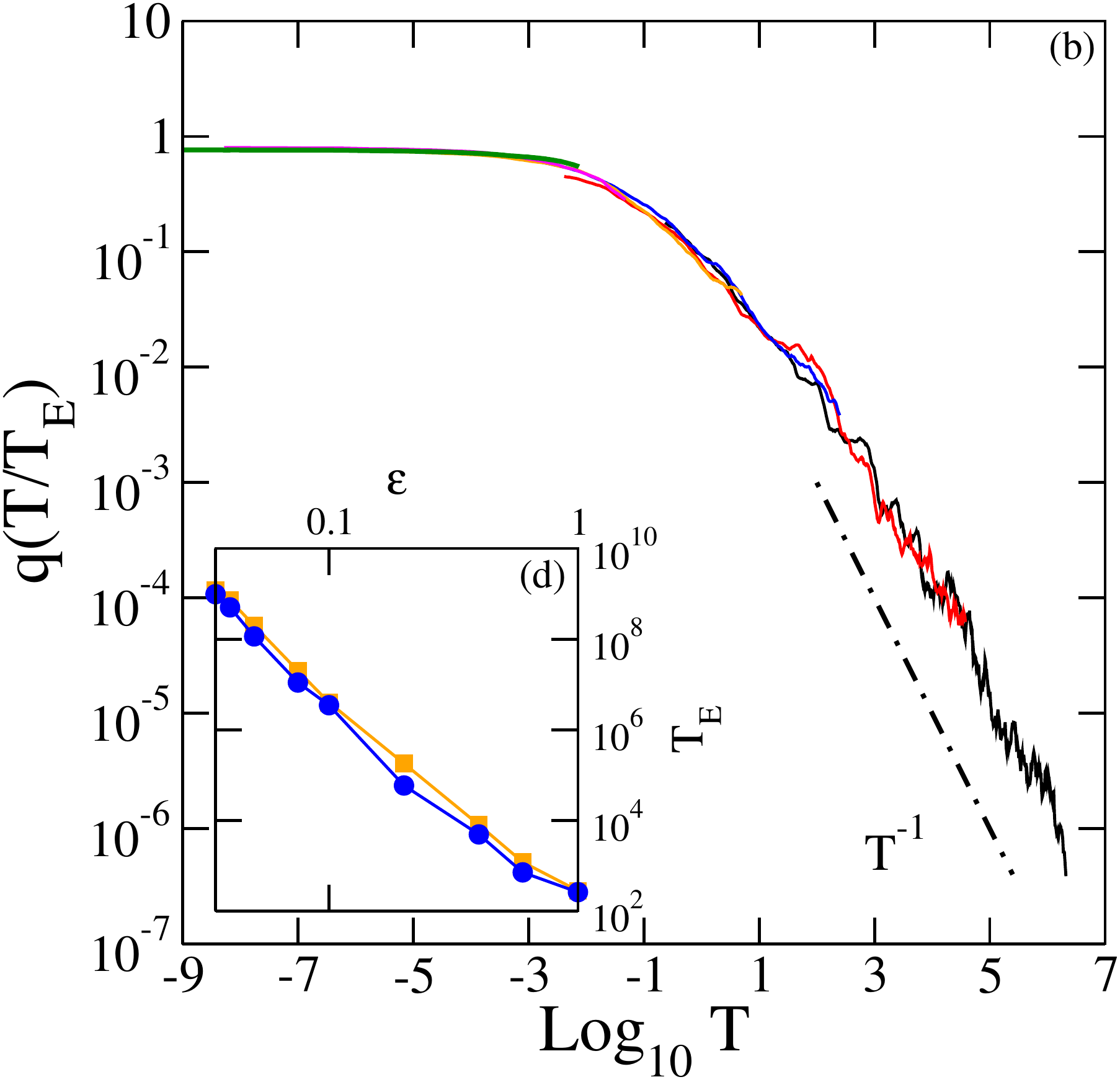}
 \caption{
 (a) $q(T/T_E)$ versus $T$ for
 the energy densities 
$h=12$ (green),
$h = 6$ (magenta), 
$h = 3$ (orange), 
$h = 0.5$ (blue),  
$h = 0.1$ (red),  
$h = 0.01$ (black)
with $\varepsilon = 0.05$. 
Inset (c): see text for details.
(b-d) Same as (a-c) versus the coupling strength
$\varepsilon=0.8$ (black),  
$\varepsilon=0.3$ (red),  
$\varepsilon=0.1$ (blue), 
$\varepsilon=0.05$ (orange), 
$\varepsilon=0.025$ (magenta), 
$\varepsilon=0.015$ (green)
with $h = 5$. 
The dashed-dotted lines guide the eye and indicate algebraic trend. 
Here $N=2^{10}$.
}
  \label{fig3_app}
\end{figure} 
In the cases shown
in Fig.~\ref{fig6}, $T_E$ was determined by first extracting the prefactor of the power-law regression of the black curve of both plots (corresponding
to $h = 0.01$ in Fig.~\ref{fig4}(a) and to $\varepsilon = 0.8$ in Fig.~\ref{fig4}(b)). 
We fix these as the two basic cases. 
Then, for all the higher $h$ or lower $\varepsilon$ cases respectively we rescaled the integration time $T\rightarrow T/x$ of the curve $q$ by a factor $x$, and select the proper $\hat{x}$ for which the rescaled curve align with their corresponding
basic cases.  
The ergodization time $T_E$ of each curve is finally obtained by multiplying its selected $\hat{x}$ with the ergodization time $T_E$ of the corresponding basic case.
In Fig.~\ref{fig3_app} we present the time-evolution of the parameter $q$ shown in Fig.~\ref{fig6} rescaled by the ergodization time $T_E$, to show the alignment
between the curves.
In the cases shown in Fig.~\ref{fig3}, the intermediate plateau exhibited by the time evolution of the index $q$ prevented us from obtaining a
proper $1/T$ fitting and the rescaling of the integration time $T$ using the techniques described above. 
Then, the ergodization time $T_E$ was determined by the introduction of a cut-off at $q=10^{-1}$ (violet dashed horizontal line in Fig.~\ref{fig1}). 
We then test the reliability of this cut-off procedure
on two SRN cases. 
In Fig.~\ref{fig3_app} we show the ergodization time $T_E$ extracted by rescaling (orange squares) and by cut-off at $q=10^{-1}$ (blue squares).
The two sets of measurements show good agreement.\\

\section{Distribution $\rho_T$ of the finite time average}\label{sec:distro_T_ave}

We here show 
the time-dependence of the distribution $\rho$, corresponding to two cases of the large energy limit $h\rightarrow \infty$ of the KG chain shown in Fig.~\ref{fig4}(b). 
\begin{figure}[h]
 \centering
 \includegraphics[ width=\columnwidth]{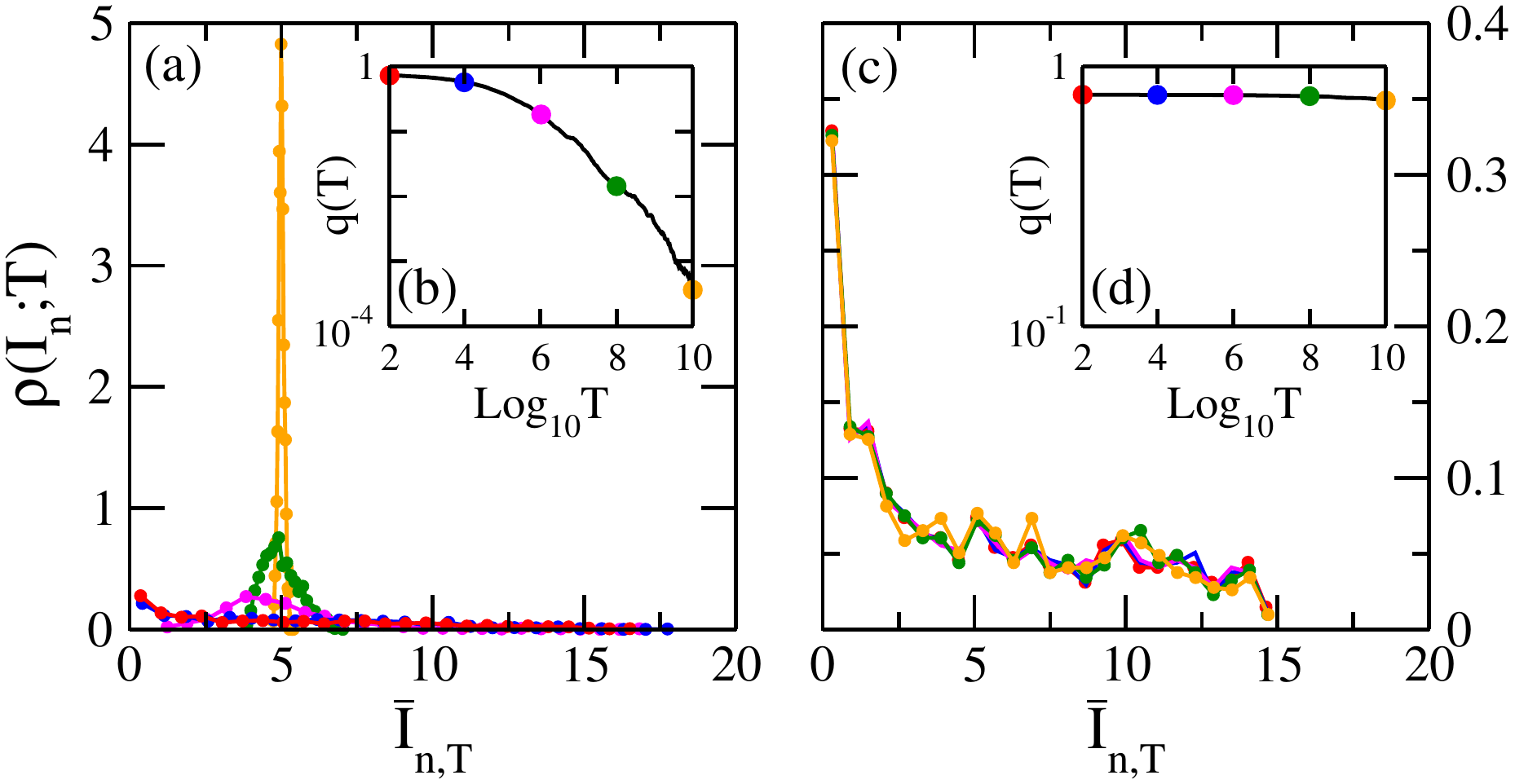}
 \caption{
 (a) $\rho(I_n;T)$ for $\varepsilon=0.1$ and different times 
$T=10^2$  (red), 
$T=10^4$    (blue), 
$T=10^6$    (magenta), 
$T=10^8$    (green), 
$T=10^{10}$ (orange)
marked in the inset (b), which reports the squared coefficient of variation $q(T)$. 
(c) and (d) Same as (a) and (b) for $\varepsilon=0.005$.
Here $h=5$, $N=2^{10}. $
}
  \label{fig4_app}
\end{figure}
Fig.~\ref{fig4_app}(a) is obtained for $\varepsilon = 0.1$. In this case, 
the decay of index $q$ corresponds a convergence of the distribution $\rho_T$ towards a delta function. In particular, between $\rho(I_n,10^6)$ (magenta) and $\rho(I_n,10^8)$ (green), the index $q$ crosses the threshold $q = 10^{-1}$ (indicated by the horizontal dashed violet line in Fig.~\ref{fig4_app}(b)) which marks the time scale $T_E \sim 3.4\cdot 10^6$. Indeed, we notice that for $T\geq T_E$, the distribution $\rho_T$ of the finite time-averages $\langle J_n\rangle_T$ begins to converge towards a delta function
Fig.~\ref{fig4_app}(b) is instead obtained for $\varepsilon = 0.005$. In this  case the index $q$ does not show any decay, and correspondently the distribution $\rho_T$ exhibits an broad trend which does not change as the integration time $T$ increases.

\section{Distributions $P_n^+$ in the SRN case}\label{sec:PDF_KG_largeE}

\begin{figure}[h]  
 \centering
 \includegraphics[ width=0.9\columnwidth]{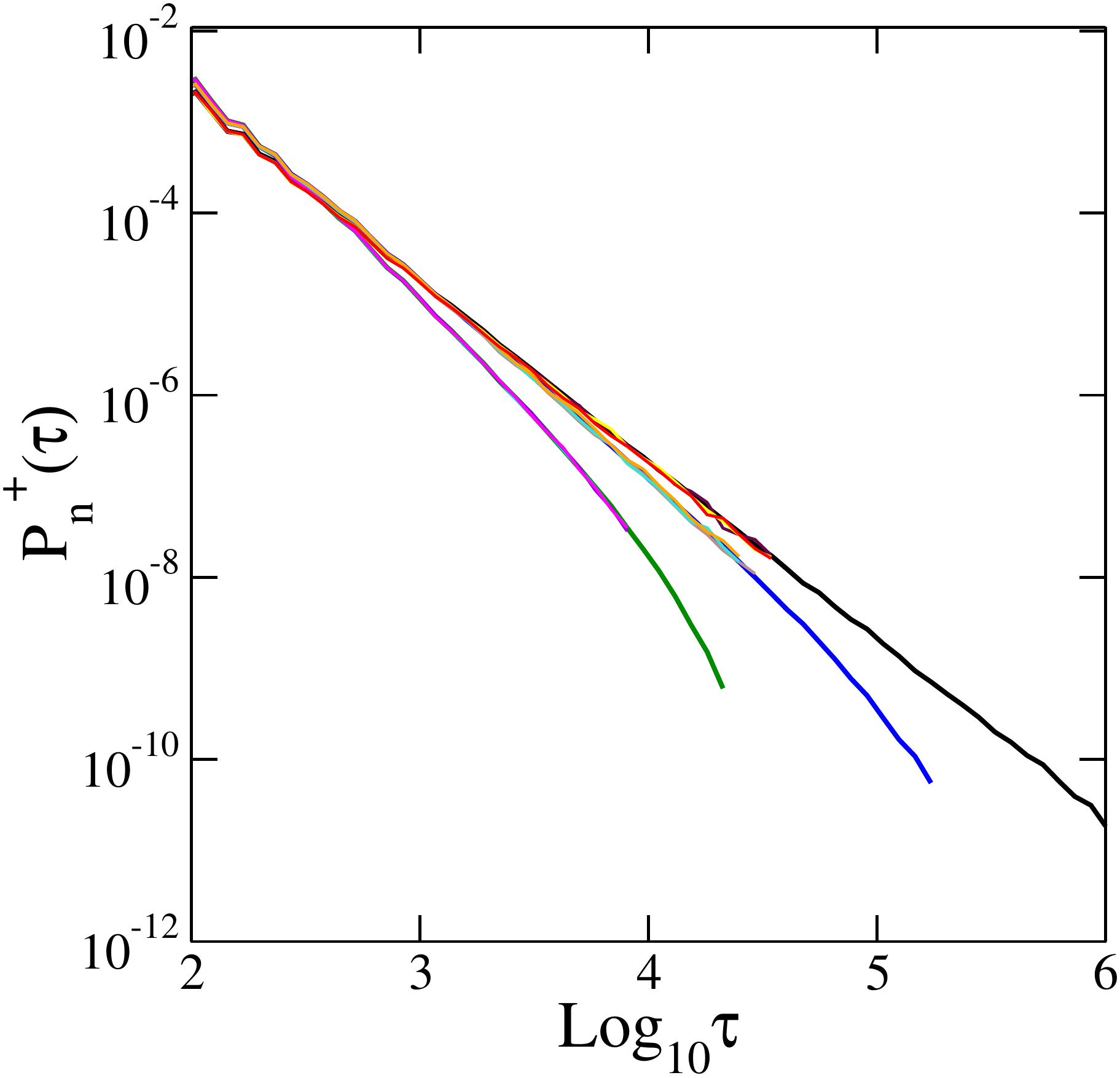}
 \caption{
 $P_n(\tau)$ obtained for $\varepsilon = 1.01$ with 
$n = 1$ (cyan), 
$n = 16$ (violet), 
$n = 32$ (magenta), 
and all the $I_n$ combined (green);
for $\varepsilon = 2.88$ with
$n = 1$ (brown), 
$n = 16$ (orange), 
$n = 32$ (turquoise), 
and all the $I_n$ combined (blue); 
for $\varepsilon = 12.9$ with 
$n = 1$  (red), 
$n = 16$ (maroon), 
$n = 32$ (yellow), 
and all the $I_n$ combined (black). 
Here the system size is $N=2^8$ and the total integration time $T=10^9$. 
}
\label{fig6_app}
\end{figure}

In Fig.~\ref{fig5} we plot the distribution function $P^+$ of the events $\tau^+$ obtained by combining the events detected by all  the observables $J_n$ in Eq.(\ref{eq:KG_observables}) in one unique distribution.
In Fig.~\ref{fig6_app} we show  
the distribution $P_n$ computed for the observable $I_1,I_{16}, I_{32}$ and the distribution $P$ for all $n$ combined for three different energies $h=1.01$ (green), $h=2.88$ (blue) and $h=12.9$ (black). 
In the three cases, the distributions  $P_n$ obtained with a single observable $I_n$ overlap with each other and with the distribution $P$ obtain by combining the events detected by all $I_n$. The curves cannot be distinguished.  
This suggests that, due to translation invariance, the excursions out of equilibrium of each action $I_n$ in Eq.(\ref{eq:KG_observables})  at a given distance from an integrable limit are governed by the same distribution.

\bibliographystyle{apsrev4}
\let\itshape\upshape
\normalem
\bibliography{weak_ergodicity}

\begin{thebibliography}{60}
\bibitem [{rat()}]{ratio_A}%
  \BibitemOpen
  \href@noop {} {\bibinfo  {journal} {We obtained the pre-factor $A$ for Fig.3
  (LRN case) and Fig.6 (SRN case) separately by computing the ratio $A \equiv
  T_E / \tau_{q}$ for one case of each figure. Specifically: for Fig.3 $A
  \equiv T_E / \tau_{q,32}^+ = 166$ is computed for mode $k=32$ at $h=0.1$; for
  Fig.6 $A \equiv T_E / \tau_{q}^+ = 132$ is computed for the weak coupling
  regime at $\varepsilon=0.1$ for $h=5$}\ }\BibitemShut {NoStop}%
\end{thebibliography}

\end{document}